\documentclass[12pt]{article}

\newcommand{\blind}{0}

\usepackage{geometry}                
\newcommand{\hide}[1]{}

\usepackage{natbib}
\usepackage{url}
\usepackage{mathtools}
\usepackage{amssymb}
\usepackage{setspace}
\usepackage{multirow}
\usepackage{threeparttable}

\hide{
\usepackage{amscd}
\usepackage{amsfonts}
\usepackage{amsmath}
\usepackage{amssymb}
\usepackage{amsthm}
\usepackage{cases}		 
\usepackage{cutwin}
\usepackage{enumerate}
\usepackage{epstopdf}
\usepackage{graphicx}
\usepackage{ifthen}
\usepackage{lipsum}
\usepackage{mathrsfs}	
\usepackage{multimedia}
\usepackage{wrapfig}
}

\newtheorem{theorem}{Theorem}
\newtheorem{lemma}{Lemma}

\newtheorem{condition}{Condition}
\newtheorem{remark}{Remark}

\newtheorem{proof}{Proof}

\def\T{\textnormal{T}}
\def\obs{\textnormal{obs}}
\def\X{\mathbf{X}}
\def\W{\mathbf{W}}
\def\V{\mathbf{V}}
\def\wald{\textnormal{wald}}
\def\ils{\textnormal{ils}}
\def\ols{\textnormal{ols}}
\def\logit{\textnormal{OB}}
\def\cal{\textnormal{COB}}
\def\var{\textnormal{var}}

\begin{document}
	\def\spacingset#1{\renewcommand{\baselinestretch}%
		{#1}\small\normalsize} \spacingset{1}

	
	\if0\blind
	{
		\title{\bf Model-assisted complier average treatment effect estimates in randomized experiments with non-compliance and a binary outcome}
		\author{Jiyang Ren\footnote{Correspondence: rjy19@mails.tsinghua.edu.cn}\\ \hspace{.2cm}
			Center for Statistical Science, Department of Industrial Engineering,\\ Tsinghua University, Beijing, 100084, China
		}
		\date{\today}
		\date{}       
		\maketitle
     
	}

	\begin{abstract}

In randomized experiments, the actual treatments received by some experimental units may differ from their treatment assignments. This non-compliance issue often occurs in clinical trials, social experiments, and the applications of randomized experiments in many other fields. Under certain assumptions, the average treatment effect for the compliers is identifiable and equal to the ratio of the intention-to-treat effects of the potential outcomes to that of the potential treatment received. To improve the estimation efficiency, we propose three model-assisted estimators
for the complier average treatment effect in randomized experiments with a binary outcome. We study their asymptotic properties, compare their efficiencies with that of the Wald estimator, and propose the Neyman-type conservative variance estimators to facilitate valid inferences. Moreover, we extend our methods and theory to estimate the multiplicative complier average treatment effect. Our analysis is randomization-based, allowing the working models to be misspecified. Finally, we conduct simulation studies to illustrate the advantages of the model-assisted methods and apply these analysis methods in a randomized experiment to evaluate the effect of academic services or incentives on academic performance.
		
	\end{abstract}
	
	\noindent%
	\vfill
	
	\newpage
	\spacingset{1.5}

\section{Introduction}
\label{sec:introduction}

Randomized experiments are widely used to discover causality in social science,  medical research, and many other fields. Under the Neyman-Rubin potential outcomes framework \citep{Neyman1923, Rubin1974}, the average treatment effect can be identified and estimated without bias under random treatment assignment and the stable unit treatment value assumption (SUTVA) \citep{Rubin:1980, imbens2015causal}.

However, in practice, some experimental units may not comply with their treatment assignments. For example, if someone is randomly assigned to take a new drug, he/she still has the right to refuse the drug and instead take the placebo. It is often difficult, or even impossible, to force experimental units to receive the assigned treatments. In this case, although the treatment assignment is randomized, the treatment received may not be.
 Non-compliance problems are common in randomized experiments in a variety of fields, such as clinical trials \citep{Hirano2000}, social experiments \citep{sherman1984}, policy evaluations \citep{schochet2003}, and educational studies \citep{james2005}. Because the experimental units are assigned to only one of the treatment arms, we cannot observe their treatment received status under the alternative treatment arm. Thus, we do not know who the compliers are, and it is difficult to interpret the estimated treatment effect. 

A highly successful application of the potential outcomes framework is to clarify the fundamental assumptions necessary to identify treatment effects in the non-compliance problems \citep{imbens1994identification,angrist1995two,angrist1996identification,imbens1997bayesian}. \citet{angrist1996identification} proposed a set of identification assumptions. Under these assumptions, the instrumental variable estimand could be interpreted as the local average treatment effect (LATE) for the subpopulation of compliers, also called the complier average treatment effect (CATE). Moreover, they showed that the CATE is equal to the ratio of intention-to-treat (ITT) effects of the potential outcomes to that of the potential treatment received. Because the ITT effects can be consistently estimated by the difference in the sample means of the potential outcomes (or potential treatment received) under the treatment and control groups, we can consistently estimate the CATE using a plug-in method, often known as the Wald estimator \citep{wald1940}.

Under a super-population framework, \citet{imbens1994identification} showed that the Wald estimator is consistent and asymptotically normal. Although the super-population framework is convenient for theoretical analysis, it is unnatural in many problems \citep{Manski2018,Abadie2020,Guo2021}. For example, when all of the states are recruited to an experiment for policy evaluation, it is not clear what the super-population refers to. In addition, if the experimental units are convenience sampled or carefully selected, they cannot be considered independent and identically distributed samples from a super-population. In these problems, the randomization-based inference is still meaningful, and the finite-population framework is more appropriate than the super-population framework. Our first contribution is to provide rigorous proof for the asymptotic normality of the Wald estimator under a finite-population and randomization-based inference framework. We assume that the potential outcomes and covariates are fixed quantities, the only source of randomness is the treatment assignment, and the treated units are sampled without replacement from the experimental units, and thus the treatment assignments are dependent.



In randomized experiments, baseline covariates are often collected for valid or more efficient inferences; however, the Wald estimator does not incorporate covariate information. To address this issue, the most popular method in econometrics is the two-stage least squares (2SLS), which 70 uses the simultaneous equations model (SEM) framework and allows for the inclusion of baseline covariates \citep{Angrist2008, Wooldridge2010}. Moreover, under a fully parametric model specification, maximum likelihood and Bayesian inferential methods are also used to estimate the CATE \citep{imbens1997bayesian, Hirano2000}. However, these methods lack validity if the working models are misspecified. Semiparametric approaches have been proposed for robust estimation \citep{Abadie2003, Tan2006, Clarke2012, Ogburn2015, Wang2021}. To further reduce the risk of model misspecification, \citet{Frolich2007} proposed a nonparametric imputation-based CATE estimator, showed its asymptotic normality, and provided a semiparametric efficiency bound. Moreover, many researchers favour inverse probability weighting methods, some of which share the doubly robust properties \citep{Tan2006, Frolich2007, Donald2014, Heiler2021, Sun2021}. All of these studies assumed that the observed data, outcomes, covariates, treatment received, and treatment assignments, are independent and identically distributed, and sampled from a super-population. In this paper, we consider a finite-population and randomization-based inference framework, propose three model-assisted CATE estimators by incorporating the covariate information, study their asymptotic properties, and compare their efficiencies with that of the Wald estimator. In our theoretical analysis, we allow the working models to be arbitrarily misspecified.

In randomized experiments with perfect compliance, researchers have proposed regression adjustment methods with treatment assignment by covariate interactions to estimate the average treatment effect, and showed that the resulting estimator is generally more efficient than the simple difference-in-means estimator \citep{lin2013,bloniarz2015lasso, fogarty2018, Yue2018, Liu2019, LiDing2020}. Motivated by these ideas and considering non-compliance, we propose an indirect least squares (ILS) method that includes the treatment assignment by covariate interactions in the two linear (working) models. We show that the proposed estimator is consistent and asymptotically normal with an asymptotic variance no larger than that of the Wald estimator.

The ILS method uses linear regression, which ignores the binary nature of the potential outcomes and treatment received. For binary outcomes, logistic regression may fit the data better. Without considering non-compliance, \citet{Guo2021} proposed a generalized Oaxaca--Blinder estimator and showed its consistency and asymptotic normality under the conditions
of prediction unbiasedness, stability, and simple prediction models. We generalize the Oaxaca--Blinder estimator to solve the non-compliance problems and establish its asymptotic theory.

We find that, although the generalized Oaxaca--Blinder estimator is consistent and asymptotically normal, it may reduce the precision compared with the Wald estimator. To address this issue, we propose a calibrated Oaxaca--Blinder estimator, borrowing an idea from \citet{Cohen2020}. We show that it is consistent, asymptotically normal, and more efficient than or at least as efficient as the Wald estimator.

We also provide conservative variance estimators to construct asymptotically conservative confidence intervals for the CATE, and extend the proposed methods to estimate and infer the multiplicative complier average treatment effect (MCATE). Finally, we evaluate the finite-sample performance of the proposed estimators through simulation studies and a real data application.

The remainder of the paper proceeds as follows: Section~\ref{sec:notation} introduces the framework, notation, and identification assumptions for the CATE in completely randomized experiments with non-compliance; Section~\ref{sec:wald} reviews the Wald estimator and establishes its finite population asymptotic normality; Section~\ref{sec:regression} describes three model-assisted CATE estimators and discusses their asymptotic properties; Section~\ref{sec:MCATE} extends the proposed methods to estimate MCATE; Section~\ref{sec:simulation} provides simulation studies, and Section~\ref{sec:realdata} analyzes a randomized experiment to evaluate the effect of academic services or incentives on academic performance; Section~\ref{sec:discussion} concludes the paper with discussions. The proofs are given in the Supplementary Material.

\section{Framework, notation, and assumptions}\label{sec:notation}

Consider a completely randomized experiment with $n$ units, of which $n_1$ units are randomly assigned to the treatment group and the remaining $n_0$ units to the control group. We denote by $Z$ the treatment assignment indicator. For unit $i$, $Z_i=1$ if it is assigned to the treatment group, and $Z_i=0$ otherwise. 
The probability distribution of the treatment assignment indicator is  $\textnormal{pr}\{(Z_1,\dots,Z_n)^{\T}=(z_1,\dots,z_n)^{\T}\}= n_1 ! n_0 ! / n!,\ z_i=0,1, \ \sum_{i=1}^n z_i = n_1$. Throughout this paper, we assume that the SUTVA holds, which allows us to conveniently define the potential outcomes.
Let $D_i(z)$ denote the binary potential treatment received status. If unit $i$ is assigned to the treatment arm $z$, $z=0,1$, $D_i(z)=1$ if it receives the active treatment and $D_i(z)=0$ if it receives the control treatment. The observed treatment received is  $D_i^{\obs}=Z_iD_i(1)+(1-Z_i)D_i(0)$. According to the joint values of the potential treatment received $(D_i(0),D_i(1))^{\T}$, the experimental units can be divided into four strata: always takers with $(1,1)^{\T}$, 
never takers with $(0,0)^{\T}$, compliers with $(0,1)^{\T}$, and defiers with $(1,0)^{\T}$.




Let $Y_i(z)$ denote the binary potential outcomes if unit $i$ is assigned to the treatment arm $z$, $z=0,1$. The observed response value is  $Y_i^{\text{obs}}=Z_iY_i(1)+(1-Z_i)Y_i(0)$. The average treatment effect of $Z$ on $Y$, also called the ITT effect, is $\tau_Y = E \{ Y_i(1)-Y_i(0) \} = \sum_{i=1}^{n}\{Y_i(1)-Y_i(0)\}/n$. Similarly, the ITT effect of $Z$ on $D$ is $\tau_D = E \{ D_i(1)-D_i(0) \} = \sum_{i=1}^{n}\{D_i(1)-D_i(0)\}/n$. 

Many studies have investigated the treatment effect of the treatment received on the response, i.e., the treatment effect of $D$ on $Y$ instead of $Z$ on $Y$. For example, if we want to evaluate the effect of a drug, we need to compare the outcomes between patients taking the drug and not taking the drug. The ITT effect $\tau_Y$ considers all of the experimental  units, including patients who always take the drug and those who never take the drug. Only for compliers does the treatment received status equal the treatment assigned status. Thus, we prefer to study the CATE, which is
$$\tau = E\{ Y_i(1)-Y_i(0) \mid D_i(1)=1, D_i(0)=0 \} =  \sum_{i \in \mathcal{C} } \{Y_i(1)-Y_i(0) \}/n_c,$$
where $ \mathcal{C} =\{i: D_i(1)=1, D_i(0)=0\}$ is the set of compliers, and $n_c$ is the total number of compliers.

As only one of the potential treatment received, $D_i(1)$ or $D_i(0)$, can be observed, we do not know who the compliers are. Thus, to determine the CATE, we use the following identification assumptions proposed by \citet{imbens1994identification} and \citet{angrist1996identification}.

\begin{condition}\label{Identification assumptions} (i) Exclusion restriction: $Y_i(1) = Y_i(0)$ for all units with $D_i(1) = D_i(0)$;  
	(ii) Monotonicity: $D_i(1)\geq D_i(0)$; (iii) Strong instrument: $\tau_D  > C_0 > 0 $, where $C_0$ is a positive constant independent of $n$.
\end{condition}

\begin{remark}
	First, the exclusion restriction implies that the treatment assignment $Z$ affects the potential outcome $Y$ only through the treatment received $D$.  Thus, for always takers and never takers with $D_i(1)=D_i(0)$, the treatment effect of $Z$ on $Y$ is 0. Second, the monotonicity assumption rules out the defiers. Third, the strong instrument assumption implies that the ITT effect of $Z$ on $D$ is not equal to 0.
\end{remark}

Under Condition \ref{Identification assumptions}, \citet{angrist1996identification} showed that the CATE is equal to the ratio of the ITT effect of $Z$ on $Y$ to that of $Z$ on $D$, denoted by
$
\tau=\tau_Y /\tau_D.
$
This identification formula has an intuitive explanation. The average treatment effect of $Z$ on $Y$ in the entire finite-population can be decomposed into the summation of the multiplication of the LATE and the proportion of its corresponding subpopulation. Under Condition \ref{Identification assumptions}, the LATE for always takers and never takers is 0, and the proportion of defiers is 0. Thus, the ITT effect of $Z$ on $Y$ is equal to the CATE multiplied by the proportion of compliers (the ITT effect of $Z$ on $D$).


To proceed, we introduce the following notation. Let $\X_i=(X_{i1},\dots,X_{ip})^{\T}$ be a $p$-dimensional vector of baseline covariates for unit $i$. As the covariates are not affected by the treatment assignment, we have $\X_i^{\obs}=\X_i(1)=\X_i(0)=\X_i$. For $R=D,Y,\X$, the  average values of the treatment received, response, and covariates in the finite-population, the treatment group, and the control group are
$$
\bar R(z)=n^{-1}\sum_{i=1}^n R_i(z), \quad
\bar R_1^{\obs}=n_1^{-1}\sum_{i=1}^n Z_iR_i^{\obs}, \quad 
\bar R_0^{\obs}=n_0^{-1}\sum_{i=1}^n (1-Z_i)R_i^{\obs}, \quad z=0,1,
$$
where $\bar\cdot$ indicates the finite-population mean, and the additional subscripts 1 and 0 indicate sample means in the treatment and control groups, respectively.

For each treatment arm $z$, $z =0, 1$, let $S_{R(z)}^2$ and $S_{R(1)-R(0)}^2$ be the finite-population variance of the potential outcomes $R(z)$ and the unit-level treatment effect $R_i(1)-R_i(0)$, respectively. Let $S_{R(z)Q(z)}$ and $S_{\{R(1)-R(0)\}\{Q(1)-Q(0)\}}$ be the finite-population covariance between the potential outcomes $R(z)$ and $Q(z)$ and between their unit-level treatment effects, respectively. We replace the uppercase $S$ with a lowercase $s$ to denote the sample analog.

\section{Wald estimator}\label{sec:wald}
As the treatment assignment is completely randomized, the ITT effects $\tau_Y$ and $\tau_D$ can be estimated without bias by the difference in the means of the observed outcomes under the treatment and control groups. The difference-in-means estimators are  $\hat\tau_{Y}=\bar{Y}_1^{\obs}-\bar{Y}_0^{\obs}$ and  $\hat\tau_{D}=\bar{D}_1^{\obs}-\bar{D}_0^{\obs}$. As $\tau=\tau_Y/\tau_D$, the Wald estimator for $\tau$ is to plug in the estimators for $\tau_Y$ and $\tau_D$. That is, $\hat\tau_{\wald}=\hat\tau_{Y}/\hat\tau_{D}$.

Under the SEM framework, which is widely used in econometrics, the Wald estimator is equivalent to the 2SLS estimator without covariates \citep{Angrist2008}. The 2SLS method by definition consists of two regression stages, i.e., fitting a linear regression of $D$ on $Z$ and then fitting a linear regression of $Y$ on the fitted values of $D$:
$$
D_i^{\obs}=\alpha_0+\alpha_1 Z_i+e_{D,i}, \quad 
Y_i^{\obs}=\beta_0+\beta_{\text{2SLS}} \hat{D}_i+e_{Y,i},
$$
where $\hat{D}_i$ is the fitted value in the first regression. The 2SLS estimator is the estimator for the second-stage regression coefficient of $\hat{D}_i$,  $\beta_{\text{2SLS}}$.
The 2SLS has been investigated extensively in the econometric literature \citep{Angrist2008,Wooldridge2010}, under a super-population framework. In what followings, we study its asymptotic properties under a finite-population and randomization-based inference framework using the finite-population central limit theorem \citep{LiDing2017}.


As $\hat\tau_{\wald}-\tau=(\hat\tau_{Y}-\tau\hat\tau_{D})/\hat\tau_D$, under mild conditions, $\hat\tau_{\wald}$ has the same asymptotic distribution as $(\hat\tau_{Y}-\tau\hat\tau_{D})/\tau_D=\hat\tau_A/\tau_D$, where $\hat\tau_A$ is the difference-in-means estimator for the ITT effect of $Z$ on the transformed potential outcomes $A$: $A_i(z)=Y_i(z)-\tau D_i(z)$, $z=0,1$. The asymptotic variance of $n^{1/2}\hat\tau_{\wald}$ is the limit of $\sigma^2_{\wald}$,  
\begin{eqnarray}
\sigma^2_{\wald} = \frac{n}{\tau_D^2} \Big\{ \frac{S_{A(1)}^2}{n_1}+\frac{S_{A(0)}^2}{n_0}- \frac{S_{A(1)-A(0)}^2}{n} \Big\}. \nonumber
\end{eqnarray}

As we cannot observe or estimate $A_i(1)$ and $A_i(0)$ simultaneously, the last term in $\sigma^2_{\wald}$ is not estimable. We directly drop it and obtain a Neyman-type  conservative variance estimator:
$
\hat \sigma^2_{\wald}=n\hat\tau_D^{-2} \{s_{A(1)}^2/n_1+s_{A(0)}^2/n_0\},
$
where
$ s_{A(z)}^2 = (n_z-1)^{-1}\sum_{i:Z_{i}=z} \{ \hat A_{i}(z) -n_z^{-1} \sum_{j:Z_j=z}  \hat A_j(z) \}^2 $ is the sample variance for the estimated outcome $\hat A_i(z) = Y_i(z) - \hat\tau_{\wald} D_i(z)$, $z = 0,1$. (A more rigorous notation for $s_{A(z)}^2$ is $s_{\hat A(z)}^2$. For ease of notation, we still use $s_{A(z)}^2$ when it does not confuse.)



To establish the asymptotic normality of $\hat \tau_{\wald}$, the following regularity
conditions are required by the finite-population central limit theorem \citep{LiDing2017}.

\begin{condition} \label{cond unadj}
	As $n\rightarrow \infty$, (i) the proportions of the treated and control units have limits between 0 and 1, i.e., $n_1/n \rightarrow p_1 \in (0,1)$ and $n_0/n \rightarrow p_0 \in (0,1)$; (ii) for $z=0,1$, the finite-population means, $\bar Y(z)$ and $\bar D(z)$, the finite-population variances, $S_{Y(z)}^2$, $S_{D(z)}^2$, $S_{Y(1)-Y(0)}^2$, and $S_{D(1)-D(0)}^2$, and the finite-population covariances, $S_{Y(z)D(z)}$ and $S_{\{Y(1)-Y(0)\} \{D(1)-D(0)\}}$, tend to finite limits, and the limit of $\sigma^2_{\wald} $ is positive.
\end{condition}

\begin{remark}
	One crucial condition for the finite-population central limit theorem is
	$
	\max_{i=1,\dots,n} \{ R_i(z) - \bar{R}(z) \}^2 / n \rightarrow 0, 
	$
	$R = Y,D$. As $Y_i(z)$ and $D_i(z)$ are binary potential outcomes, they are bounded and clearly satisfy this condition. 
\end{remark}



\begin{theorem}\label{thm::unadj}
	Under Conditions \ref{Identification assumptions} and \ref{cond unadj}, $\hat\tau_{\wald}-\tau$ converges in probability to 0 and $n^{1/2}(\hat\tau_{\wald}-\tau)/\sigma_{\wald}$ converges in distribution to $N(0,1)$.
	Furthermore, $\hat \sigma^2_{\wald}$ converges in probability to the limit of 
	$
	n\tau_D^{-2}\{S_{A(1)}^2/n_1+S_{A(0)}^2/n_0 \},
	$
	which is no less than that of $\sigma^2_{\wald}$.
\end{theorem}


Theorem \ref{thm::unadj} provides a normal approximation for the distribution of $\hat\tau_{\wald}$. The variance estimator is generally conservative. It is consistent if and only if the unit-level treatment effect of $A_i(z)$ is constant, i.e., $Y_i(1) - Y_i(0) -  \tau \{ D_i(1) - D_i(0) \} = C$ for some constant $C$.
Based on Theorem~\ref{thm::unadj}, an asymptotically conservative confidence interval for $\tau$ is $[\hat\tau_{\wald}-q_{\alpha/2}n^{-1/2}\hat\sigma_{\wald},
\hat\tau_{\wald}+q_{\alpha/2}n^{-1/2}\hat\sigma_{\wald}]$, where $\alpha \in (0,1)$ is the significance level and $q_{\alpha/2}$ is the upper $\alpha/2$ quantile of a standard normal distribution. The asymptotic coverage rate of the above confidence interval is greater than or equal to $1 - \alpha$.


\section{Model-assisted CATE estimators}\label{sec:regression}
Although the Wald estimator exhibits good asymptotic properties, it does not use covariate information. If these baseline covariates can predict the potential outcomes, covariate adjustment tends to reduce the variance of the estimated treatment effect \citep{lin2013,bloniarz2015lasso, fogarty2018, Yue2018, Liu2019, LiDing2020}. 
In this section, we propose three model-assisted CATE estimators and study their asymptotic properties under the finite-population and randomization-based inference framework. The working models used to obtain these estimators can be arbitrarily misspecified.

\subsection{Indirect Least Squares estimator with interactions}
\label{subsec:ils}

In econometrics, the ILS and 2SLS are two widely used methods to estimate the CATE under the SEM framework \citep{Hayashi2000,Angrist2008,Wooldridge2010}. The ILS regresses the observed potential outcomes $Y_i^\obs$ and $D_i^\obs$ on the treatment assigned status $Z_i$ and covariates $\X_i$:
$$
D_i^\obs=\alpha_0+\alpha_1 Z_i+\X_i^{\T}\alpha_2+e_{D,i},
\quad
Y_i^\obs=\beta_0+\beta_1 Z_i+\X_i^{\T}\beta_2+e_{Y,i}.
$$
The ILS estimator estimates the ratio of two regression coefficients of $Z_i$, $\beta_{\ils}=\beta_1/\alpha_1$ \citep{Angrist2008}. The 2SLS method, using information from the covariates, performs the following two regressions:
$$
D_i^\obs=\alpha_0+\alpha_1 Z_i+\X_i^{\T}\alpha_2+e_{D,i}, \quad 
Y_i^\obs=\beta_0+\beta_{\text{2SLS}} \hat{D}_i+\X_i^{\T}\beta_2+e_{Y,i},
$$
where $\hat{D}_i$ is the fitted value in the first regression. 
The 2SLS estimator estimates the second stage regression coefficient of $\hat D_i$, $\beta_{\text{2SLS}}=\text{Cov}(Y,\tilde{Z})/\text{Cov}(D,\tilde{Z})$, 
where $\tilde{Z}$ is the residual from a regression of $Z$ on $\X$. The 2SLS estimator with a single instrument $Z$ is identical to the ILS estimator \citep{Angrist2008}.


Considering a randomization-based inference without imposing the linear modelling assumptions on the potential outcomes, neither regression in the ILS estimator can ensure efficiency gains relative to the estimator without adjusting the covariates \citep{Freedman2008}. For example, in the second regression, the OLS estimator for $\beta_1$ may have a larger asymptotic variance than that of the unadjusted difference-in-means estimator $\hat\tau_{Y}$ for estimating $\tau_Y$. Thus, the ILS estimator may not always be more efficient than the Wald estimator $\hat\tau_{\wald}$. To address this issue, we can add the treatment assignment by covariate interactions in both ILS regressions, utilizing the idea of \cite{lin2013}. Intuitively, if we can estimate $\tau_Y$ and $\tau_D$ more accurately, we can improve the efficiency of the CATE estimator.

Adding interactions is equivalent to regressing $Y_i^{\obs}$ and $D_i^{\obs}$ on $\X_i$ in the treatment and control groups separately. The OLS estimators for the slopes are
$$
\hat{\beta}_{R}(z)=\underset{\beta\in\mathbb{R}^{p}}{\arg \min } \sum_{i: Z_i=z} \left\{R_{i}^{\obs}-\bar{R}_z^{\obs}-(\X_i-\bar{\X}_z^{\obs})^{\T} \beta\right\}^{2},\quad R=Y,D, \quad z=0,1.
$$
The OLS estimators for $\tau_Y$ and $\tau_D$ are 
$$
\hat\tau_{\ols,R}=\Big\{\bar{R}_1^{\obs}-(\bar{\X}_{1}^{\obs}-\bar{\X})^{\T} \hat{\beta}_{R}(1)\Big\}-\Big\{\bar{R}_0^{\obs}-(\bar{\X}_0^{\obs}-\bar{\X})^{\T} \hat{\beta}_{R}(0)\Big\},\quad R=Y,D.
$$
Then, we can obtain the ILS estimator with interactions: $\hat\tau_{\ils}=\hat\tau_{\ols,Y}/\hat\tau_{\ols,D}$.

To study the asymptotic properties of $\hat\tau_{\ils}$, we first define the projection coefficients:
$$
\beta_{R}(z)=\underset{\beta\in\mathbb{R}^{p}}{\arg \min } \sum_{i=1}^n \Big\{R_{i}(z)-\bar{R}(z)-(\X_i-\bar{\X})^{\T} \beta\Big\}^{2},\quad R=Y,D, \quad z=0,1.
$$
Then, we  decompose the potential outcomes $Y_i(z)$ and $D_i(z)$ into two parts, i.e., projection on the space spanned by the linear combination of the covariates and projection error:
$$
R_i(z)=\bar R(z)+(\X_i-\bar{\X})^{\T}\beta_R(z)+\epsilon_{R,i}(z),\quad R=Y,D, \quad z=0,1.
$$
The above equations are only used to define the projection errors $\epsilon_{Y,i}(z)$ and $\epsilon_{D,i}(z)$, and all quantities in these equations are fixed.


Similar to the Wald estimator, we define the transformed potential outcomes $A_{\ols}$:
$$
A_{\ols,i}(z)=Y_i(z)-\tau D_i(z)-(\X_i-\bar{\X})^{\T}\left\{\beta_Y(z)-\tau \beta_D(z) \right\}, \quad
z=0,1. $$
As shown in the following Theorem~\ref{thm::ols}, the asymptotic variance of $n^{1/2}\hat\tau_{\ils}$ is the limit of $\sigma^2_{\ils}$,
\begin{eqnarray}
\sigma^2_{\ils}=\frac{n}{\tau_D^2} \Big\{ \frac{S_{A_{\ols}(1)}^2}{n_1}+\frac{S_{A_{\ols}(0)}^2}{n_0}- \frac{S_{A_{\ols}(1)-A_{\ols}(0)}^2}{n} \Big\} \nonumber.
\end{eqnarray}
Similar to the $\sigma^2_{\wald}$ estimation,  we drop the last term to derive a conservative variance estimator. Specifically, we define 
$$
\hat A_{\ols,i}(z)=Y_i(z)-\hat\tau_{\ils}D_i(z)-(\X_i-\bar{\X})^{\T}\{\hat{\beta}_Y(z)-\hat\tau_{\ils} \hat{\beta}_D(z) \}
$$ 
by plugging in the unknown quantities $\tau$, $\beta_Y(z)$, and $\beta_D(z)$.  Then, we use the sample variance of $\hat A_{\ols}(z)$ to estimate $S_{A_{\ols}(z)}^2$, denoted by
$
s_{A_{\ols}(z)}^2=(n_z-p-1)^{-1}\sum_{i:Z_i=z}\{\hat A_{\ols,i}(z)-n_z^{-1}\sum_{j:Z_j=z}\hat A_{\ols,j}(z)\}^2,
$
where the factor $(n_z-p-1)^{-1}$ adjusts for the degrees of freedom to achieve a better finite-sample performance.
Then, a conservative estimator for $\sigma^2_{\ils}$ is
$
\hat\sigma^2_{\ils}=n\hat\tau_{\ols,D}^{-2} \{ s_{A_{\ols}(1)}^2/n_1+s_{A_{\ols}(0)}^2/n_0\}.
$



To establish the asymptotic normality of $\hat \tau_{\ils}$, we need the following regularity condition:
\begin{condition}\label{cond ols}
	As $n\rightarrow \infty$, (i) for each covariate $X_k$ $(k = 1, ..., p)$, $
	\max_{i\in \{1,...,n\}}\{X_{ik}-\bar{X}_{k}\}^2/n \rightarrow 0$;
	(ii) for $z=0,1$, the finite-population covariances, $S_{\X }^2$, $S_{\X Y(z)}$, and  $S_{\X D(z)}$, tend to finite limits with $S_{\X}^2$ and its limit being strictly positive definite, and the limit of $\sigma^2_{\ils}$ is positive.
\end{condition}


\begin{theorem}\label{thm::ols}
	Under Conditions~\ref{Identification assumptions}--\ref{cond ols}, $\hat\tau_{\ils}-\tau$ converges in probability to 0 and $n^{1/2}(\hat\tau_{\ils}-\tau)/\sigma_{\ils}$ converges in distribution to $N(0,1)$. Furthermore, $\hat\sigma^2_{\ils}$ converges in probability to the limit of 
	$
	n\tau_D^{-2}\{ S_{A_{\ols}(1)}^2/n_1+S_{A_{\ols}(0)}^2/n_0\},
	$
	which is no less than that of $\sigma_{\ils}^2$.
\end{theorem}

Theorem~\ref{thm::ols} provides a normal approximation for the distribution of $\hat\tau_{\ils}$ under complete randomization. The variance estimator is generally conservative. It is consistent if and only if the unit-level treatment effect of $A_{\ols,i}(z)$ is constant, i.e., $A_{\ols,i}(1)-A_{\ols,i}(0)=C$ for some constant $C$.
Based on Theorem~\ref{thm::ols}, an asymptotically conservative confidence interval for $\tau$ is $[\hat\tau_{\ils}-q_{\alpha/2}n^{-1/2}\hat\sigma_{\ils},
\hat\tau_{\ils}+q_{\alpha/2}n^{-1/2}\hat\sigma_{\ils}]$, whose asymptotic coverage rate is greater than or equal to $1-\alpha$. Comparing the asymptotic variances and variance estimators of $\hat\tau_{\ils}$ and  $\hat\tau_{\wald}$, we have the following result:


\begin{theorem}\label{thm::ols var}
	Under the conditions of Theorems~\ref{thm::unadj} and \ref{thm::ols},
	the difference between the asymptotic variances of $n^{1/2}\hat\tau_{\ils}$ and $n^{1/2}\hat\tau_{\wald}$ is the limit of
	$
	-( \tau_D^2  p_{1} p_{0})^{-1}  \{p_{0} \delta(1)+p_{1} \delta(0)\}^{\T} S_{\X}^2\{p_{0} \delta(1)+p_{1} \delta(0)\} \leq 0,
	$
	and the difference between the variance estimators, $\hat\sigma^2_{\ils}$ and $\hat\sigma^2_{\wald}$, converges in probability to the limit of
	$
	-n\tau_D^{-2}\{\delta(1)^{\T} S_{\X}^2 \delta(1)/n_{1}+\delta(0)^{\T}  S_{\X}^2  \delta(0)/n_{0}\}\leq 0,
	$
	where $\delta(1)=\beta_Y(1)-\tau\beta_D(1)$ and  $\delta(0)=\beta_Y(0)-\tau\beta_D(0)$.
\end{theorem}

Theorem~\ref{thm::ols var} shows that both the asymptotic variance and variance estimator of $\hat\tau_{\ils}$ are no greater than those of $\hat\tau_{\wald}$. Thus, the ILS estimator with interactions improves or at least does not degrade the estimation and inference efficiencies. 
The improvement in asymptotic efficiency mainly depends on the $R^2$ of the projection of $Y_i(z)-\tau D_i(z)$ on $\X_i$, which measures the variance of the potential outcomes that is explained by the covariates. In particular, if $\delta(1)=\delta(0)=0$, i.e., $\X_i$ does not affect $Y_i(z)-\tau D_i(z)$,  $\hat\tau_{\ils}$ has no asymptotic efficiency gain compared with $\hat\tau_{\wald}$. For another special case with $p_{0} \delta(0)+p_{1} \delta(0)=0$, although $\hat\tau_{\ils}$ has no asymptotic efficiency improvement, it can produce a shorter confidence interval because of the variance estimator, if at least one of $\delta(1)$ and  $\delta(0)$ is not equal to 0.

\subsection{Logistic Oaxaca--Blinder estimator}
\label{subsec:logit}

The ILS estimator uses a linear regression model to fit the data. However, for a binary outcome,
it is natural to consider a logistic regression model to improve the efficiency further. To motivate the method, we discuss an imputation-based interpretation of the OLS estimator $\hat\tau_{\ols,Y}$ \citep{imbens2015causal,Guo2021}. We can derive $\hat\tau_{\ols,Y}$ by the following imputation procedure: we fit a linear regression model of $Y_i^{\obs}$ on $(1, \X_i^{\T})^{\T}$ using the data in the treatment group. Thereafter, for any unit $i$ in the control group, we impute the unobserved potential outcome $Y_i(1)$, denoted by $\hat Y_i(1)$, by the fitted model and obtain $\hat \tau_{Y,i} = \hat Y_i(1) - Y_i(0) $. Similarly, we can obtain $\hat \tau_{Y,i}$ for unit $i$ in the treatment group. As shown by \cite{imbens2015causal} and  \cite{Guo2021}, $\hat\tau_{\ols,Y} = n^{-1} \sum_{i=1}^{n} \hat \tau_{Y,i}$. In econometrics, this double-imputation procedure is known as the Oaxaca--Blinder method \citep{Blinder1973, Oaxaca1973,Kline2011}.

We can use nonlinear models, such as logistic regression, Poisson regression, or smoothing splines, to impute the unobserved potential outcomes if they fit the data better than the linear model and obtain a generalized Oaxaca--Blinder estimator \citep{Guo2021}. We extend this idea to the non-compliance case. In this paper, $Y$ and $D$ are binary, and thus we consider using the logistic regression models to fit the data and impute the unobserved potential outcomes.

For $z=0,1$, to fill in the unobserved potential outcome $Y_i(z)$ for units assigned to the treatment arm $1-z$, we use data from the treatment arm $z$ and fit logistic regression models:
$$
\hat{\theta}_{Y}(z)=\underset{\theta\in\mathbb{R}^{p+1}}{\arg\min}\sum_{i: Z_i=z}\Big\{-Y_i^{\obs} \tilde{\X}_i^{\T} \theta+\log \Big(1+e^{ \tilde{\X}_i^{\T}\theta}\Big)\Big\},
\quad
\hat{\mu}_{Y,i}(z)=\frac{e^{\tilde{\X}_i^{\T}\hat{\theta}_{Y}(z)}}{1+e^{\tilde{\X}_i^{\T}\hat{\theta}_{Y}(z)}},
$$
where $\tilde{\X}_i=(1,\X_i^{\T})^{\T}$, and $\hat{\mu}_{Y,i}(z)$ is the predicted probability of $Y_i(z)=1$. 
Then, we fill in the unobserved values of $Y_{i}(1)$ and $Y_{i}(0)$ using 
$$
\hat{Y}_{i}(1)=\Bigg\{\begin{array}{ll}Y_{i}(1)  &Z_{i}=1 \\ \hat{\mu}_{Y,i}(1) &Z_{i}=0
\end{array} ,
\quad 
\hat{Y}_{i}(0)=\Bigg\{\begin{array}{ll}\hat{\mu}_{Y,i}(0) &Z_{i}=1 \\ Y_{i}(0)  &Z_{i}=0
\end{array},
$$
and obtain the estimator for $\tau_Y$: $\hat\tau_{\logit,Y}= n^{-1} \sum_{i=1}^n\{\hat{Y}_{i}(1)-\hat{Y}_{i}(0)\}$. 
Similarly, we can obtain $\hat\tau_{\logit,D}= n^{-1} \sum_{i=1}^n\{\hat D_i(1)-\hat D_i(0)\}$.
Then, the logistic Oaxaca--Blinder estimator for $\tau$ is 
$\hat{\tau}_{\logit}=\hat\tau_{\logit,Y}/\hat\tau_{\logit,D}$.

\begin{remark}
	When $\tilde{\X}$ is rank-deficient, or the 1's and 0's among the observed outcomes in the treatment or control group can be perfectly separated by a hyperplane, the logistic regression coefficient estimators, $\hat{\theta}_{Y}(z)$ or $\hat{\theta}_{D}(z)$, may not be unique or even exist. In this case, variable selection or regularization is necessary to obtain these estimators.
\end{remark}

To investigate the asymptotic properties of $\hat{\tau}_{\logit}$, we define $\theta_{Y}(z)$ and $\theta_{D}(z)$ as the solutions of the following finite-population logistic regression problems: for $R=Y,D$, $z=0,1$,
$$
\theta_{R}(z)= \underset{\theta\in \mathbb{R}^{p+1}}{\arg\min}\left[ L_{R}^{(n)}(\theta):L_{R}^{(n)}(\theta)=\frac{1}{n} \sum_{i=1}^{n}\left\{-R_i(z) \tilde{\X}_{i}^{\T} \theta+\log \left(1+e^{\tilde{\X}_{i}^{\T} \theta}\right)\right\}\right].
$$
Throughout the paper, we assume that $\theta_{Y}(z)$ and $\theta_{D}(z)$ exist and are unique. 
The predicted probabilities are 
$
\mu_{R,i}(z)={\exp\{ \tilde{\X}_i^{\T}\theta_{R}(z) \}} / [ {1+\exp\{\tilde{\X}_i^{\T}\theta_{R}(z)\}} ],
$
and the residuals are $\eta_{R,i}(z)=R_i(z)-\mu_{R,i}(z)$, $R=Y,D$, $z=0,1$, where $\theta_{R}(z)$, $\mu_{R,i}(z)$ and $\eta_{R,i}(z)$ all depend on $n$. For ease of notation, we omit this dependence when it does not confuse.  

We define the transformed potential outcomes $A_{\logit}$ as follows: 
$$
A_{\logit,i}(z)=Y_i(z)-\tau D_i(z)-\{\mu_{Y,i}(z)-\tau \mu_{D,i}(z)\}=\eta_{Y,i}(z)-\tau \eta_{D,i}(z), \quad
z=0,1. $$
As shown in Theorem~\ref{thm::logit}, the asymptotic variance of $n^{1/2}\hat{\tau}_{\logit}$ is the limit of $\sigma^2_{\logit}$,  
$$
\sigma^2_{\logit}=\frac{n}{\tau_D^2}\Big\{\frac{S_{A_{\logit}(1)}^2}{n_1}+\frac{S_{A_{\logit}(0)}^2}{n_0}-  \frac{S_{A_{\logit}(1)-A_{\logit}(0)}^2}{n} \Big\}.
$$
Similar to the $ \sigma^2_{\ils}$ estimation, we can obtain a conservative variance estimator:  $\hat\sigma^2_{\logit}=n\hat\tau_{\logit,D}^{-2}\{s_{A_{\logit}(1)}^2/n_1+s_{A_{\logit}(0)}^2/n_0\}$,
where $s_{A_{\logit}(z)}^2=(n_z-p-1)^{-1}\sum_{i: Z_i=z}\{ \hat A_{\logit,i}(z) - n_z^{-1} \sum_{j: Z_j=z} \hat A_{\logit,j}(z) \}^2$ is the sample variance of the estimated outcomes 
$$\hat A_{\logit,i}(z)= Y_i(z)- \hat\tau_{\logit} D_i(z)-\{\hat\mu_{Y,i}(z)-\hat\tau_{\logit}\hat\mu_{D,i}(z)\}, \quad z=0,1.$$



We require the following condition to derive the asymptotic normality of $\hat\tau_{\logit}$:



\begin{condition}\label{cond logit}
	(i) For all large $n$, $R=Y,D$, and $z=0,1$, $\nabla^{2} L_{R}^{(n)} \{\theta_{R}(z) \} - C I_{(p+1) \times (p+1)}$ is positive semi-definite for some constant $C>0$ independent of $n$, where $I_{(p+1) \times (p+1)}$ is a $(p+1) \times (p+1)$ identity matrix; 
	(ii) the fourth moments of the covariates are uniformly bounded, i.e., $n^{-1} \sum_{i=1}^{n} X_{ik}^4 \leq M$ for $k=1,\dots ,p$ and some constant $M < \infty$;
	(iii) as $n\rightarrow \infty$, for $z=0,1$, the finite-population variances, $S^2_{\eta_{Y}(z)}$, $S^2_{\eta_{D}(z)}$, $S^2_{\eta_{Y}(1)-\eta_{Y}(0)}$, and $S^2_{\eta_{D}(1)-\eta_{D}(0)}$, and the finite-population covariances, $S_{\eta_{Y}(z)\eta_{D}(z)}$ and $S_{\{\eta_{Y}(1)-\eta_{Y}(0)\}\{\eta_{D}(1)-\eta_{D}(0)\}}$,
	tend to finite limits, and the limit of $\sigma^2_{\logit}$ is positive.
\end{condition}



\begin{theorem}\label{thm::logit}
	Under Conditions~\ref{Identification assumptions}, \ref{cond unadj} and \ref{cond logit}, $\hat\tau_{\logit}-\tau$ converges in probability to 0 and $n^{1/2}(\hat\tau_{\logit}-\tau)/\sigma_{\logit}$ converges in distribution to $N(0,1)$. Furthermore,  $\hat\sigma^2_{\logit}$ converges in probability to the limit of $n\tau_{D}^{-2}\{S_{A_{\logit}(1)}^2/n_1+S_{A_{\logit}(0)}^2/n_0\}$, which is no less than that of $\sigma^2_{\logit}$.
\end{theorem}

Theorem \ref{thm::logit} provides a normal approximation for the distribution of $\hat\tau_{\logit}$. Again, the variance estimator is generally conservative, and is consistent if and only if the unit-level treatment effect $A_{\logit,i}(1)-A_{\logit,i}(0)$ is constant. Based on Theorem \ref{thm::logit}, we can obtain an asymptotically conservative confidence interval for $\tau$: $[\hat\tau_{\logit}-q_{\alpha/2}n^{-1/2}\hat\sigma_{\logit},
\hat\tau_{\logit}+q_{\alpha/2}n^{-1/2}\hat\sigma_{\logit}]$. However, $\hat\tau_{\logit,Y}$ and $\hat\tau_{\logit,D}$ cannot always improve the asymptotic efficiency compared with the difference-in-means estimators $\hat\tau_{Y}$ and $\hat\tau_{D}$. We refer to \citet{Cohen2020} for a counterexample. Thus,  $\hat\tau_{\logit}$ may degrade the efficiency in some extreme cases compared with $\hat\tau_{\wald}$. 



\subsection{Calibrated Oaxaca--Blinder estimator}
\label{subsec:cal}
Although the logistic regression model might be more appropriate to predict binary potential outcomes than the linear regression model,  $\hat\tau_{\logit}$ cannot ensure efficiency gains. To solve this non-inferiority problem, we propose a calibrated Oaxaca--Blinder estimator, borrowing techniques from \citet{Cohen2020}.


The basic idea is to take the logistic regression fitted values $\hat{\mu}_{Y,i}(z)$ and $\hat{\mu}_{D,i}(z)$ as new covariates ${\W}_{i}^{\obs}=(\hat{\mu}_{Y,i}(1), \hat{\mu}_{Y,i}(0),\hat{\mu}_{D,i}(1),\hat{\mu}_{D,i}(0))^{\T}$, and regress $Y_i^{\obs}$ and $D_i^{\obs}$ on ${\W}_{i}^{\obs}$ 
in the treatment and control groups separately. The coefficient estimators are
$$
\hat{\gamma}_{R}(z)=\underset{\gamma\in\mathbb{R}^4}{\arg\min } \sum_{i: Z_i=z}\Big\{R_{i}^{\obs}-\bar{R}_z^{\obs}-({\W}_{i}^{\obs}-\bar{{\W}}_z^{\obs})^{\T}  \gamma\Big\}^{2}, \quad R=Y,D, \quad z=0,1.
$$
We obtain the estimators for $\tau_Y$ and $\tau_D$:
$$
\hat\tau_{\cal,R}=\Big\{\bar{R}_1^{\obs}-(\bar{{\W}}_{1}^{\obs}-\bar{{\W}}^{\obs})^{\T} \hat{\gamma}_{R}(1)\Big\}-\Big\{\bar{R}_0^{\obs}-(\bar{{\W}}_{0}^{\obs}-\bar{{\W}}^{\obs})^{\T} \hat{\gamma}_{R}(0)\Big\}, \quad R=Y,D,
$$
where $\bar{{\W}}^{\obs}=n^{-1}\sum_{i=1}^n{\W}_{i}^{\obs}$. Then, the calibrated Oaxaca--Blinder estimator for $\tau$ is $\hat\tau_{\cal}=\hat\tau_{\cal,Y}/\hat\tau_{\cal,D}$. 

\begin{remark}
	In practice, if some of the potential outcomes are sparse (the proportion of elements 1 is close to zero), the logistic regression model might overfit the data such that the fitted probabilities are numerically 0 or 1. Over-fitting makes the fitted model lack robustness. Any slight disturbance significantly affects the prediction result and may destabilize the calibrated linear regressions. When faced with this issue, we should carefully use the calibrated Oaxaca--Blinder estimator, or more robustly, not perform the calibration step.
\end{remark}

To study the asymptotic properties of $\hat\tau_{\cal}$,  we decompose the potential outcomes $Y_i(z)$ and $D_i(z)$ into  projections on the space spanned by the linear combination of the covariates $\W_i=(\mu_{Y,i}(1), \mu_{Y,i}(0),\mu_{D,i}(1),\mu_{D,i}(0))^{\T}$ and projection errors:
$$
R_i(z)=\bar R(z)+(\W_i-\bar{\W})^{\T}\gamma_R(z)+\xi_{R,i}(z),\quad R = Y,D, \quad z = 0, 1,
$$
where $\gamma_{R}(z)$ is the finite-population projection coefficient, 
$$
\gamma_{R}(z)=\underset{\gamma\in\mathbb{R}^{4}}{\arg \min } \sum_{i=1}^n \Big\{R_{i}(z)-\bar{R}(z)-(\W_i-\bar{\W})^{\T} \gamma\Big\}^{2}, \quad R = Y,D,
$$
and $\xi_{R,i}(z)$ is the projection error.

We define the transformed potential outcomes $A_{\cal}$ as follows: 
$$
A_{\cal,i}(z)=Y_i(z)-\tau D_i(z)-(\W_i-\bar{\W})^{\T}\left\{\gamma_Y(z)-\tau \gamma_D(z) \right\}, \quad z=0,1.$$
The asymptotic variance of $n^{1/2}\hat\tau_{\cal}$ is  the limit of $\sigma^2_{\cal}$,
$$
\sigma^2_{\cal}=\frac{n}{\tau_D^2}\Big\{\frac{S_{A_{\cal}(1)}^2}{n_1}+\frac{S_{A_{\cal}(0)}^2}{n_0}- \frac{S_{A_{\cal}(1)-A_{\cal}(0)}^2}{n} \Big\}.
$$
Similar to the $\sigma_{\ils}^2$ estimation, let $\hat A_{\cal,i}(z)$ be the estimated value of $A_{\cal,i}(z)$:
$$
\hat A_{\cal,i}(z)=Y_i(z)-\hat\tau_{\cal} D_i(z)-(\W_i^{\obs}-\bar{\W}^{\obs})^{\T}\left\{\hat\gamma_Y(z)-\hat\tau_{\cal} \hat\gamma_D(z) \right\},
$$ and let $
s_{A_{\cal}(z)}^2=(n_z-5)^{-1}\sum_{i: Z_i=z}\{\hat A_{\cal,i}(z)-n_z^{-1}\sum_{j: Z_j=z}\hat A_{\cal,j}(z)\}^{2}
$
be the sample variance of $\hat A_{\cal,i}(z)$ under treatment arm $z$. Then, $\sigma^2_{\cal}$ can be conservatively estimated by 
$
\hat\sigma^2_{\cal}=n\hat\tau_{\cal,D}^{-2}\{s_{A_{\cal}(1)}^2/n_1+s_{A_{\cal}(0)}^2/n_0\}.
$

\begin{condition}\label{cond cal}
	
	As $n\rightarrow \infty$, for $z=0,1$, the finite-population covariances, $S_{\W}^2$, $S_{\W Y(z)}$, and $S_{\W D(z)}$, tend to finite limits with $S_{\W}^2$ and its limit being strictly positive definite, and the limit of $\sigma^2_{\cal}$ is positive.
\end{condition}

\begin{theorem}\label{thm::cal}
	Under Conditions~\ref{Identification assumptions}, \ref{cond unadj}, \ref{cond logit}, and \ref{cond cal}, $\hat\tau_{\cal}-\tau$ converges in probability to 0 and $n^{1/2}(\hat\tau_{\cal}-\tau)/\sigma_{\cal}$ converges in distribution to $N(0,1)$. Furthermore, $\hat\sigma^2_{\cal}$ converges in probability to the limit of
	$
	n\tau_D^{-2}\{S_{A_{\cal}(1)}^2/n_1+S_{A_{\cal}(0)}^2/n_0\},
	$
	which is no less than that of $\sigma^2_{\cal}$.
\end{theorem}

Theorem~\ref{thm::cal} provides a normal approximation for the distribution of $\hat\tau_{\cal}$, which can be used to construct an asymptotically conservative confidence interval for $\tau$: 
$[\hat\tau_{\cal}-q_{\alpha/2}n^{-1/2}\hat\sigma_{\cal},
\hat\tau_{\cal}+q_{\alpha/2}n^{-1/2}\hat\sigma_{\cal}]$.
Moreover, both the asymptotic variance and the variance estimator of $\hat\tau_{\cal}$ are no greater than those of $\hat\tau_{\wald}$ and $\hat\tau_{\logit}$, as indicated by the following Theorem:

\begin{theorem}\label{thm::cal var}
	Under the conditions of Theorems~\ref{thm::unadj}, \ref{thm::ols}, \ref{thm::logit}, and \ref{thm::cal},
	the difference between the asymptotic variances of  $n^{1/2}\hat\tau_{\cal}$ and $n^{1/2}\hat\tau_{\wald}$ is the limit of 
	$
	-(\tau_D^2p_{1} p_{0})^{-1} \{p_{0} \phi(1)+p_{1} \phi(0)\}^{\T} S_{\W}^2\{p_{0} \phi
	(1)+p_{1} \phi(0)\}\leq 0;
	$
	the difference between the variance estimators $\hat\sigma_{\cal}^2$ and $\hat\sigma^2_{\wald}$ converges in probability to the limit of 
	$
	-n\tau_D^{-2}\{\phi(1)^{\T} S_{\W}^2 \phi(1)/n_{1}+\phi(0)^{\T}  S_{\W}^2  \phi(0)/n_{0}\}\leq 0,
	$
	where $\phi(1)=\gamma_Y(1)-\tau\gamma_D(1)$ and  $\phi(0)=\gamma_Y(0)-\tau\gamma_D(0)$; and the asymptotic variance and variance estimator of $\hat\tau_{\cal}$ are less than or equal to those of $\hat\tau_{\logit}$ (the conclusion for the variance estimator holds in probability).
\end{theorem}

Theorem~\ref{thm::cal var} shows that $\hat\tau_{\cal}$ generally outperforms  $\hat\tau_{\wald}$ and $\hat\tau_{\logit}$.
If $p_{0} \phi(1)+p_{1} \phi(0) = 0$ but $\phi(1) \neq 0$ (or $\phi(0) \neq 0$),  $\hat\tau_{\cal}$ has the same asymptotic distribution as $\hat\tau_{\wald}$, while its variance estimator is generally smaller than that of $\hat\tau_{\wald}$, resulting in a narrower confidence interval. If $\phi(1)=\phi(0)=0$, $\hat\tau_{\cal}$ has no asymptotic efficiency improvement and has the same confidence interval length as $\hat\tau_{\wald}$. 
The relative asymptotic efficiency of $\hat\tau_{\cal}$ and $\hat\tau_{\ils}$ depends on which models, logistic regression or linear regression, fit the data better.

\section{Extension to multiplicative complier average treatment effect }
\label{sec:MCATE}
The CATE is a natural choice to evaluate the complier treatment effect. When $Y$ is binary, another interesting and well-studied estimand is the MCATE \citep{Didelez2010, Clarke2012, Ogburn2015, Wang2021}, defined as the treatment relative risk:
$$
\tau_M=\frac{E\left\{Y_{i}(1)\mid D_{i}(0)=0, D_{i}(1)=1\right\}}{E\left\{Y_{i}(0) \mid D_{i}(0)=0, D_{i}(1)=1\right\}}.
$$

The MCATE is well-defined when we assume that $E\left\{Y_{i}(0) \mid D_{i}(0)=0, D_{i}(1)=1\right\}\not=0$ almost surely. Let $G_i(z)=Y_i(z)D_i(z)$ and $H_i(z)=Y_i(z)\{1-D_i(z)\}$.   \citet{Abadie2002} showed that under Condition \ref{Identification assumptions}, MCATE can be identified as
$$
\tau_M=-\frac{E\left\{Y_i(1)D_i(1)\right\}-E\left\{Y_i(0)D_i(0)\right\}}{E\left[Y_i(1)\left\{1-D_i(1)\right\}\right]-E\left[Y_i(0)\left\{1-D_i(0)\right\}\right]}=-\frac{\tau_G}{\tau_H},
$$
where $\tau_{G}$ is the ITT effect of $Z$ on $YD$, and $\tau_H$ is the ITT effect of $Z$ on $Y(1-D)$. We consider $G_i(z)$ and $H_i(z)$ as new binary potential outcomes, and use the methods  introduced in Sections~\ref{sec:wald} and \ref{sec:regression} to estimate $\tau_{G}$ and $\tau_{H}$ separately. Then, we take a ratio (plug-in) to estimate $\tau_M$: $\hat{\tau}_M=-\hat{\tau}_{G}/\hat{\tau}_{H}$. All of the theoretical results on the CATE estimation can be easily extended to the MCATE estimation. The detailed asymptotic results  are presented in the Supplementary Material.

\begin{remark}
	$Y_i(z)\{1-D_i(z)\}$ is the product of two binary potential outcomes, which may make $\tau_{H}$ close to 0. In this case, all of the MCATE estimators may be unstable. In practice, we should be cautious when $\tau_M$ is the target estimand.
\end{remark}


\section{Simulation}\label{sec:simulation}
In this section, we evaluate the finite-sample performance of the Wald estimator and three model-assisted estimators through a simulation study. We consider both the CATE and MCATE. The treatment received and potential outcomes are generated by the following models:
$$
D_i(0)=I(2X_{i1}-1>0), \quad D_i(1)=I(2X_{i1}+3>0),
$$
$$
y_{i0} \mid \X_{i} \sim \text{Bernoulli}\Big(\frac{e^{-3X_{i2}}}{1+e^{-3X_{i2}}}\Big),
\quad 
y_{i1} \mid \X_{i} \sim \text{Bernoulli}\Big(\frac{e^{-3X_{i2}+1}}{1+e^{-3X_{i2}+1}}\Big),
$$
\begin{equation}
Y_{i}(0)=\Bigg\{\begin{array}{l}
y_{i0} \quad \text{if } D_i(0)=0\\
y_{i1} \quad \text{if } D_i(0)=1\\
\end{array},
\quad
Y_{i}(1)=\Bigg\{\begin{array}{l}
y_{i0} \quad \text{if } D_i(1)=0\\
y_{i1} \quad \text{if } D_i(1)=1\\
\end{array},\nonumber
\end{equation}
where $\X_i$ follows a two-dimensional normal distribution with mean zero and covariance $\Sigma$, which entries $\Sigma_{11}=\Sigma_{22}=2$ and $\Sigma_{12}=\Sigma_{21}= \rho$.

The potential outcomes and covariates are generated once and then kept fixed. 
We set $n=200, 500$,  $n_1/ n =0.3, 0.4, 0.5$, and $\rho =0,1$. The proportions of compliers, always takers, and never takers are approximately 0.5, 0.36, and 0.14, respectively. We perform completely randomized experiments 1000 times to compare the performance of $\hat\tau_{\wald}$, $\hat\tau_{\ils}$, $\hat\tau_{\logit}$, and $\hat\tau_{\cal}$, in terms of bias, standard deviation (SD), root mean square error (RMSE), empirical coverage probability (CP), and mean confidence interval length (CI length) of the $95\%$ confidence intervals (CI).


\begin{figure}
	\centering
	\includegraphics[width=0.9\linewidth]{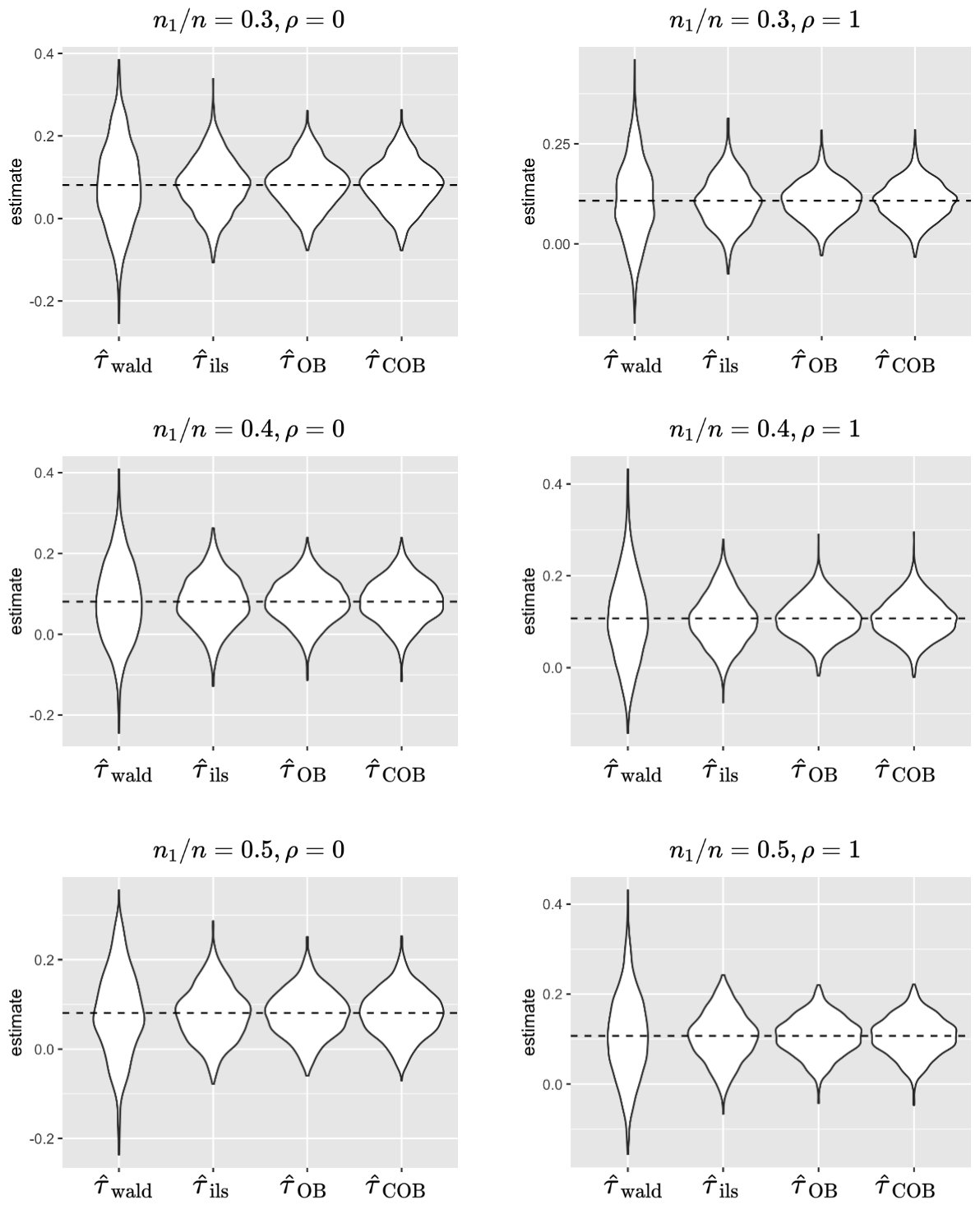}
	\caption{\label{fig::CATE500}Violin plot for the distributions of CATE estimators ($n=500$). The dotted line is the true value of the CATE.}
\end{figure}

\begin{table}
	\centering
	\caption{\label{tab::CATE500}Performance of  CATE estimators ($n=500$)}
	\begin{threeparttable}
		\begin{tabular}{cccccccccc}
	\hline
	&\multirow{2}{*}{$\rho$}& \multirow{2}{*}{$n_1/n$} & \multirow{2}{*}{Bias} & \multirow{2}{*}{SD} & \multirow{2}{*}{RMSE} & RMSE & \multirow{2}{*}{CP} & CI & Length \\ 
	& & &  &  &  & ratio &  & length & ratio\\ 
	\hline
	$\hat\tau_{\wald}$ &0& 0.3 & 0.000 & 0.102 & 0.102 & 1.000 & 0.955 & 0.407 & 1.000 \\  
	$\hat\tau_{\ils}$ & 0&0.3  & 0.001 & 0.066 & 0.066 & 0.643 & 0.965 & 0.274 & 0.673 \\ 
	$\hat\tau_{\logit}$ &0& 0.3  & 0.001 & 0.057 & 0.057 & 0.557 & 0.962 & 0.239 & 0.586 \\  
	$\hat\tau_{\cal}$ &0& 0.3  & 0.001 & 0.057 & 0.057 & 0.561 & 0.962 & 0.238 & 0.585 \\ 
	\hline
	$\hat\tau_{\wald}$ &0& 0.4  & 0.000 & 0.098 & 0.098 & 1.000 & 0.960 & 0.381 & 1.000 \\ 
	$\hat\tau_{\ils}$ &0& 0.4  & -0.000 & 0.063 & 0.063 & 0.645 & 0.956 & 0.257 & 0.675 \\ 
	$\hat\tau_{\logit}$ &0& 0.4  & 0.000 & 0.054 & 0.054 & 0.546 & 0.966 & 0.225 & 0.591 \\ 
	$\hat\tau_{\cal}$ &0& 0.4  & -0.000 & 0.054 & 0.054 & 0.547 & 0.966 & 0.224 & 0.588 \\ 
	\hline
	$\hat\tau_{\wald}$ &0& 0.5  & 0.000 & 0.095 & 0.095 & 1.000 & 0.964 & 0.375 & 1.000 \\  
	$\hat\tau_{\ils}$ &0& 0.5  & -0.001 & 0.059 & 0.059 & 0.629 & 0.971 & 0.253 & 0.674 \\  
	$\hat\tau_{\logit}$ &0& 0.5  & -0.001 & 0.052 & 0.052 & 0.553 & 0.967 & 0.221 & 0.590 \\   
	$\hat\tau_{\cal}$ &0& 0.5& -0.001 & 0.052 & 0.052 & 0.553 & 0.961 & 0.220 & 0.588 \\ 
	\hline
	$\hat\tau_{\wald}$ &1& 0.3  & 0.000 & 0.098 & 0.098 & 1.000 & 0.963 & 0.405 & 1.000 \\   
	$\hat\tau_{\ils}$ &1& 0.3  & -0.001 & 0.060 & 0.060 & 0.610 & 0.967 & 0.259 & 0.640 \\ 
	$\hat\tau_{\logit}$ &1& 0.3  & -0.000 & 0.047 & 0.047 & 0.476 & 0.971 & 0.210 & 0.520 \\  
	$\hat\tau_{\cal}$ &1& 0.3  & -0.000 & 0.047 & 0.047 & 0.481 & 0.968 & 0.210 & 0.518 \\ 
	\hline
	$\hat\tau_{\wald}$ &1& 0.4 & 0.001 & 0.093 & 0.093 & 1.000 & 0.965 & 0.379 & 1.000 \\   
	$\hat\tau_{\ils}$ &1& 0.4  & -0.002 & 0.054 & 0.055 & 0.585 & 0.980 & 0.243 & 0.641 \\ 
	$\hat\tau_{\logit}$ &1& 0.4  & 0.000 & 0.043 & 0.043 & 0.459 & 0.976 & 0.198 & 0.523 \\ 
	$\hat\tau_{\cal}$ &1& 0.4  & 0.000 & 0.043 & 0.043 & 0.462 & 0.975 & 0.198 & 0.521 \\ 
	\hline
	$\hat\tau_{\wald}$ &1& 0.5  & -0.001 & 0.092 & 0.092 & 1.000 & 0.960 & 0.373 & 1.000 \\  
	$\hat\tau_{\ils}$ &1& 0.5  & -0.002 & 0.055 & 0.055 & 0.594 & 0.983 & 0.239 & 0.640 \\ 
	$\hat\tau_{\logit}$ &1& 0.5  & -0.002 & 0.043 & 0.043 & 0.463 & 0.978 & 0.195 & 0.522 \\   
	$\hat\tau_{\cal}$ &1& 0.5 & -0.002 & 0.043 & 0.043 & 0.466 & 0.979 & 0.194 & 0.520 \\ 
	\hline
\end{tabular}

		Note: SD, standard deviation; RMSE, root of mean squared error; RMSE ratio, relative to the Wald  estimator; CP, empirical coverage probability of the $95\%$ confidence intervals; CI length, mean confidence interval length of the $95\%$ confidence intervals; Length ratio, relative to the Wald estimator. 
	\end{threeparttable}
	
\end{table}

The results for CATE  are shown in Figure~\ref{fig::CATE500} and Table~\ref{tab::CATE500} for $n=500$, and the results for $n=200$ and the MCATE are presented in the Supplementary Material. From these results,
we observe that, first, the biases of all of the methods are negligible in accordance with their
asymptotic unbiasedness. Second, compared with $\hat\tau_{\wald}$, $\hat\tau_{\ils}$ reduces the RMSE and CI length by $35.5\%-41.5\%$ and $32.5\%-36.0\%$,  respectively. The logistic regression based estimator $\hat\tau_{\logit}$ further reduces the RMSE and CI length by $7.6\%-13.4\%$ and $8.4\%-12.4\%$, respectively. The calibrated estimator $\hat\tau_{\cal}$ performs similarly to $\hat\tau_{\logit}$. 
Furthermore, the empirical coverage probabilities of the $95\%$ confidence intervals produced by $\hat\tau_{\wald}$, $\hat\tau_{\ils}$, $\hat\tau_{\logit}$ and $\hat\tau_{\cal}$ are higher than $95\%$ in all of the cases, because of the conservative variance estimators. 

\section{Real data Analysis}\label{sec:realdata}
In this section, we apply the proposed methods to analyze a real data set from the Student Achievement and Retention Project (STAR), a randomized trial to evaluate the effect of academic services or incentives on academic performance among first-year college students \citep{Angrist2009}. This trial was conducted at one of the satellite campuses of a large Canadian university. All of the first-year students entering in September 2005 were randomly assigned to the treatment or control group. The students in the treatment group were offered academic support services, financial incentives, or a combination of services and incentives, and they were required to sign up for consent; otherwise, they were ineligible for services and incentives. $Z_i$ denotes whether the student $i$ was assigned to the treatment group, and $D_i$ denotes whether the student $i$ signed up for consent. The monotonicity assumption seems reasonable because the students assigned to the control group were unlikely to sign up for consent. There were $n = 1461$ students in total, and the cross-tabulation of the treatment assigned and  received is shown in Table~\ref{tab::cross-tabulation}.

\begin{table}[!h]
	\centering
	\caption{\label{tab::cross-tabulation}Number of the treatment assigned and received}
			\begin{threeparttable}
	\begin{tabular}{c|cc}
	\hline 
	&  Sign up ($D=1$) & No sign up ($D=0$)\\
	\hline 
	Treatment ($Z=1$)& 409 & 157 \\ 
	Control ($Z=0$) & 0 & 895 \\ 
	\hline 
\end{tabular}
		\end{threeparttable}
\end{table}

The outcome of interest is whether the students were in good standing after one year. To evaluate the effect of services or incentives on college achievement, we estimate both the CATE and MCATE using the following 11 covariates to perform regression adjustment (linear or logistic): gender, age, high school GPA, whether mother tongue is English, whether lives at home, whether at first-choice school, whether plans to work while in school, whether rarely or never puts off studying for tests, whether wants more than a BA, whether intends to finish in 4 years, and parents’ education.

The point estimates and $95\%$ CIs are presented in Tables~\ref{tab::real-data}. The results show that the support of services or incentives has no significant effect on students' good standing after one year. This conclusion is the same as that drawn by \citet{Angrist2009}. Moreover, compared with $\hat\tau_{\wald}$, all of the model-assisted methods, $\hat\tau_{\ils}$, $\hat\tau_{\logit}$, and $\hat\tau_{\cal}$, reduce the variance by, $10.6\%$, $10.9\%$, $12.3\%$ for the CATE, and $14.2\%$, $14.6\%$, $16.0\%$ for the MCATE, respectively, and thus produce shorter confidence intervals.

\begin{table}[!h]
	\centering
	\caption{\label{tab::real-data}
		Results for the STAR data}
	
			\begin{threeparttable}
	\begin{tabular}{ccccc}
	\hline
	Effect&Method & Point estimator & $95\%$ CI & Variance reduction \\ 
	\hline
	\multirow{4}{*}{CATE}&$\hat\tau_{\wald}$ & 0.050 & [-0.022,0.123] & 0 \\
	&$\hat\tau_{\ils}$ & 0.044 & [-0.025,0.113]& 0.106\\ 
	&$\hat\tau_{\logit}$ & 0.044 & [-0.025,0.112] & 0.109\\ 
	&$\hat\tau_{\cal}$ & 0.044 & [-0.024,0.112] & 0.123 \\
	\hline
	\multirow{4}{*}{MCATE}&$\hat\tau_{\wald}$ & 1.108&[0.942,1.273]&0\\ 
	&$\hat\tau_{\ils}$ & 1.093&[0.940,1.246]&0.142\\ 
	&$\hat\tau_{\logit}$ & 1.091&[0.938,1.244]&0.146\\ 
	&$\hat\tau_{\cal}$ &  1.091 & [0.939,1.242] & 0.160 \\
	\hline
\end{tabular}	
	
	Note: CI, confidence interval; Variance reduction, relative to the Wald estimator.
			\end{threeparttable}
\end{table}





\begin{figure}[!h]
	\centering
	\includegraphics[width=1\linewidth]{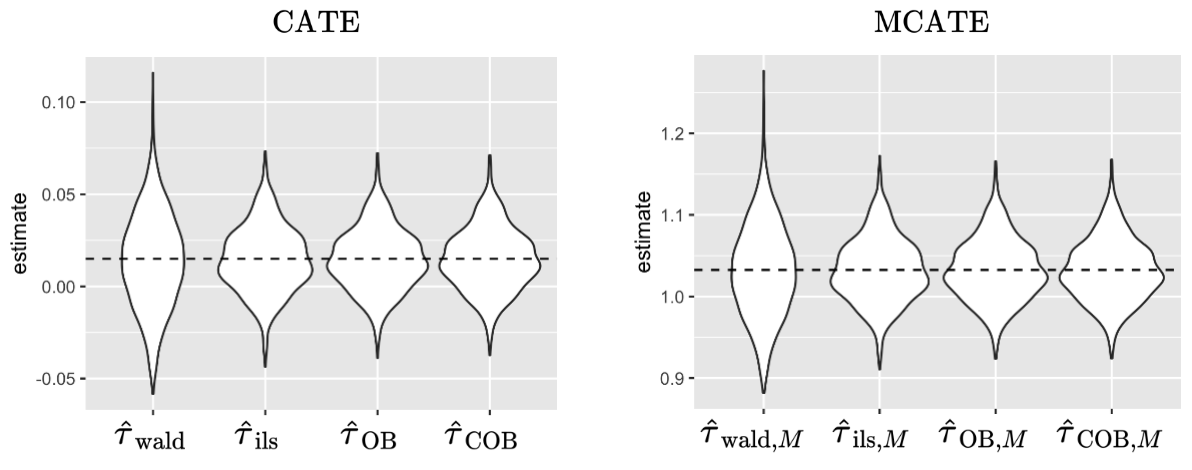}
	\caption{\label{fig::real-data-sim}Violin plot for the distributions of CATE (or MCATE) estimators in real data simulation. The dotted line is the true value of the CATE (or MCATE).}
\end{figure}

As we do not observe all of the potential outcomes, we cannot know the true gains of the model-assisted methods. For this purpose, we impute all of the missing potential outcomes using fitted logistic regression models under the monotonicity constraint. For these synthetic data, the true values of the CATE, MCATE, and the proportion of compliers are 0.015, 1.032, and 0.872, respectively. We perform completely randomized experiments 1000 times to evaluate the performance of different methods. The results are shown in Figure~\ref{fig::real-data-sim} and Table~\ref{tab::real-data-sim}. 

The conclusions are similar to those in the simulation section. First, the biases of all the methods are negligible. Second, compared with $\hat\tau_{\wald}$, $\hat\tau_{\ils}$, $\hat\tau_{\logit}$, and $\hat\tau_{\cal}$ decrease RMSE by $30\% - 35\%$ and CI length by $20\%- 25\%$, and thus they improve efficiency. Among these methods, the model-assisted method based on logistic regression perform better than that based on linear regression, and the calibrated logistic regression method performs the best. Finally, the empirical coverage probabilities of $95\%$ confidence intervals are all higher than $95\%$ because of conservative variance estimators.

\begin{table}[!h]
	\centering
	\caption{\label{tab::real-data-sim}Performance of CATE and MCATE estimators in real data simulation}
			\begin{threeparttable}
	\begin{tabular}{ccccccccc}
	\hline
	\multirow{2}{*}{Effect}&\multirow{2}{*}{Method} & \multirow{2}{*}{Bias} & \multirow{2}{*}{SD} & \multirow{2}{*}{RMSE} & RMSE& \multirow{2}{*}{CP} & \multirow{2}{*}{CI length} & Length\\ 
	&&&&& ratio &&& ratio \\ 
	\hline
	\multirow{4}{*}{CATE}&$\hat\tau_{\wald}$ & -0.001 & 0.026 & 0.026 & 1.000 & 0.982 & 0.120 & 1.000 \\ 
	&$\hat\tau_{\ils}$ & -0.000 & 0.018 & 0.018 & 0.690 & 0.989 & 0.094 & 0.786 \\ 
	&$\hat\tau_{\logit}$& -0.000 & 0.017 & 0.017 & 0.654 & 0.989 & 0.090 & 0.753 \\ 
	&$\hat\tau_{\cal}$ & -0.000 & 0.017 & 0.017 & 0.652 & 0.989 & 0.090 & 0.747 \\ 
	\hline
	\multirow{4}{*}{MCATE}&$\hat\tau_{\wald}$& -0.001 & 0.057 & 0.057 & 1.000 & 0.976 & 0.266 & 1.000 \\ 
	&$\hat\tau_{\ils}$& 0.000 & 0.040 & 0.040 & 0.691 & 0.987 & 0.209 & 0.787 \\ 
	&$\hat\tau_{\logit}$ & -0.000 & 0.038 & 0.038 & 0.664 & 0.991 & 0.203 & 0.763 \\ 
	&$\hat\tau_{\cal}$ & -0.001 & 0.038 & 0.038 & 0.663 & 0.992 & 0.201 & 0.755 \\ 
	\hline
\end{tabular}

Note: SD, standard deviation; RMSE, root of mean squared error; RMSE ratio, relative to the Wald  estimator; CP, empirical coverage probability of the $95\%$ confidence intervals; CI length, mean confidence interval length of the $95\%$ confidence intervals; Length ratio, relative to the Wald estimator. 
		\end{threeparttable}
\end{table}

\section{Discussion}\label{sec:discussion}
In this paper, we study how to efficiently estimate the CATE and MCATE in a completely randomized experiment with non-compliance and a binary outcome. 

Under the finite-population framework and mild conditions, we proved that the Wald estimator is consistent and asymptotically normal, and the variance estimator is generally conservative. 

To improve the estimation efficiency, we proposed three model-assisted methods, the ILS estimator with interactions, logistic Oaxaca--Blinder estimator, and  calibrated Oaxaca--Blinder estimator, and established their asymptotic theories. Our analysis is purely randomization-based, allowing the working model to be  misspecified. We showed that the ILS estimator with interactions is no worse than the Wald estimator;  
the logistic Oaxaca--Blinder estimator cannot ensure efficiency gains relative to the Wald estimator; the calibrated Oaxaca--Blinder estimator is generally more efficient than the Wald estimator and the logistic Oaxaca--Blinder estimator. The efficiencies of the calibrated Oaxaca--Blinder estimator and the ILS estimator with interactions are not ordered unambiguously, depending on which models, linear or logistic, fit the data better. 
In addition, 
we proposed conservative variance estimators to facilitate inferences. 


This paper focuses on estimating the CATE and MCATE in completely randomized experiments with non-compliance and binary outcomes. Other causal estimands, such as the complier odds ratio, are also interesting and worthy of further investigation. The identification of the complier odds ratio is more complex than the CATE and MCATE, and is left to future research.  
Moreover, regression adjustment methods have been widely used to improve the estimation efficiency in more complicated randomized experiments, such as stratified randomized experiments, paired randomized experiments, and cluster randomized experiments \citep{Liu2019, fogarty2018, su2021cluster}. It would also be interesting to extend the proposed methods to estimate the treatment effect in these experiments when non-compliance problems occur.


\bibliographystyle{rss} 
\bibliography{causal}       

\newpage

\appendix
\setcounter{equation}{0}
\renewcommand{\theequation}{S.\arabic{equation}}
\setcounter{figure}{0}
\renewcommand{\thefigure}{S.\arabic{figure}}
\setcounter{table}{0}
\renewcommand{\thetable}{S.\arabic{table}}
\setcounter{prop}{0}
\renewcommand{\theprop}{S.\arabic{prop}}
\setcounter{assumption}{0}
\renewcommand{\theassumption}{S.\arabic{assumption}}
\setcounter{theorem}{0}
\renewcommand{\thetheorem}{S.\arabic{theorem}}
\setcounter{condition}{0}
\renewcommand{\thecondition}{S.\arabic{condition}}
\setcounter{remark}{0}
\renewcommand{\theremark}{S.\arabic{remark}}

\begin{center}
\LARGE {\textbf{Supplementary Material}}
\end{center}
\bigskip

\section{Theoretical results for MCATE estimators}

Let us denote the Wald, ILS with interactions, logistic Oaxaca--Blinder, and calibrated Oaxaca--Blinder point and variance estimators for MCATE by $(\hat\tau_{\wald,M}, \hat\sigma^2_{\wald,M})$, $(\hat\tau_{\ils,M}, \hat\sigma^2_{\ils,M})$, $(\hat\tau_{\logit,M}, \hat\sigma^2_{\logit,M})$, and $(\hat\tau_{\cal,M}, \hat\sigma^2_{\cal,M})$, respectively. Let $\xrightarrow{d}$ and $\xrightarrow{p}$ denote convergence in distribution and in probability, respectively.

Similar to the study of the asymptotic properties of the CATE estimators, 
we define the following transformed potential outcomes: for $z=0,1$,
$$
B_i(z)=G_i(z)-\tau_M H_i(z),
$$
$$
B_{\ols,i}(z)=G_i(z)-\tau_M H_i(z)-(\X_i-\bar{\X})^{\T}\left\{\beta_G(z)-\tau_M \beta_H(z) \right\},
$$
$$
B_{\logit,i}(z)=\eta_{G,i}(z)-\tau_M \eta_{H,i}(z),
$$
$$
B_{\cal,i}(z)=G_i(z)-\tau_M H_i(z)-(\V_i-\bar{\V})^{\T}\left\{\gamma_G(z)-\tau_M \gamma_H(z) \right\},
$$
where $\beta_G(z)$, $\beta_H(z)$, $\eta_{G,i}(z)$, $\eta_{H,i}(z)$, $\V_i$, $\gamma_G(z)$, and $\gamma_H(z)$ are defined similarly to $\beta_Y(z)$, $\beta_D(z)$, $\eta_{Y,i}(z)$, $\eta_{D,i}(z)$, $\W_i$, $\gamma_Y(z)$, and $\gamma_D(z)$.

Let us denote
$$
\sigma^2_{\wald,M}= \frac{n}{\tau_H^2}\Big\{\frac{S_{B(1)}^2}{n_1}+\frac{S_{B(0)}^2}{n_0}- \frac{S_{B(1)-B(0)}^2}{n} \Big\},
$$
$$
\sigma^2_{\ils,M}  =\frac{n}{\tau_H^2} \Big\{\frac{S_{B_{\ols}(1)}^2}{n_1}+\frac{S_{B_{\ols}(0)}^2}{n_0}- \frac{S_{B_{\ols}(1)-B_{\ols}(0)}^2}{n} \Big\},
$$
$$
\sigma^2_{\logit,M}=\frac{n}{\tau_H^2}\Big\{\frac{S_{B_{\logit}(1)}^2}{n_1}+\frac{S_{B_{\logit}(0)}^2}{n_0}- \frac{S_{B_{\logit}(1)-B_{\logit}(0)}^2}{n} \Big\},
$$
$$
\sigma^2_{\cal,M}=\frac{n}{\tau_H^2}\Big\{\frac{S_{B_{\cal}(1)}^2}{n_1}+\frac{S_{B_{\cal}(0)}^2}{n_0}-\frac{S_{B_{\cal}(1)-B_{\cal}(0)}^2}{n} \Big\}.
$$

To study the asymptotic properties of the MCATE estimators, we need the following condition. Recall that, for $R = G,H$ and $z=0,1$,
$$
\theta_{R}(z)=\underset{\theta\in \mathbb{R}^{p+1}}{\arg\min}\left[ L_{R}^{(n)}(\theta):L_{R}^{(n)}(\theta)=\frac{1}{n} \sum_{i=1}^{n}\left\{-R_i(z) \tilde{\X}_{i}^{\T} \theta+\log \left(1+e^{\tilde{\X}_{i}^{\T} \theta}\right)\right\}\right].
$$	

\begin{condition}\label{cond MCATE}
	As $n\rightarrow \infty$, for $z=0,1$,
	
	(1) the proportions of the treated and control units have limits between 0 and 1, i.e., $n_1/n \rightarrow p_1 \in (0,1)$ and $n_0/n \rightarrow p_0 \in (0,1)$;
	
	(2) the finite-population means, $\bar G(z)$ and $\bar H(z)$, the finite-population variances, $S_{G(z)}^2$, $S_{H(z)}^2$, $S_{G(1)-G(0)}^2$, and $S_{H(1)-H(0)}^2$, and the finite-population covariances, $S_{G(z)H(z)}$ and $S_{\{G(1)-G(0)\} \{H(1)-H(0)\}}$, tend to finite limits, and the limit of $\sigma^2_{\wald,M}$  is positive;
	
	(3)  for each covariate $X_{k}$ $(k = 1, ..., p)$, $
	\max_{i\in \{1,...,n\}}\{X_{ik}-\bar{X}_{k}\}^2/n \rightarrow 0$;
	
	(4) the finite-population covariances, $S_{\X}^2$, $S_{\X G(z)}$, and $S_{\X H(z)}$, tend to finite limits with $S_{\X}^2$ and its limit being strictly positive definite, and the limit of $\sigma^2_{\ils,M}$ is positive;
	
	(5) for all large $n$ and $R=G,H$, $\nabla^{2} L_{R}^{(n)}\{\theta_{R}(z)\} - C I_{(p+1) \times (p+1)}$ is positive semidefinite for some constant $C>0$ independent of $n$, and
	the fourth moments of the covariates $\X_i$ are uniformly bounded;
	
	(6) the finite-population variances, $S^2_{\eta_{G}(z)}$, $S^2_{\eta_{H}(z)}$, $S^2_{\eta_{G}(1)-\eta_{G}(0)}$, and $S^2_{\eta_{H}(1)-\eta_{H}(0)}$, and the finite-population covariances, $S_{\eta_{G}(z)\eta_{H}(z)}$ and $S_{\{\eta_{G}(1)-\eta_{G}(0)\}\{\eta_{H}(1)-\eta_{H}(0)\}}$, tend to finite limits, and the limit of $\sigma^2_{\logit,M}$ is positive;
	
	(7) the finite-population covariances, $S_{\V}^2$, $S_{\V G(z)}$, and $S_{\V H(z)}$, tend to finite limits with $S_{\V }^2$ and its limit being strictly positive definite, and the limit of $\sigma^2_{\cal,M}$ is positive.
\end{condition}

\begin{theorem}\label{thm::MCATE} Suppose that Condition \ref{Identification assumptions} holds.
	
	(1) Under Condition~\ref{cond MCATE} (1)--(2),  $\hat\tau_{\wald,M}-\tau_M\xrightarrow{p}0$ and $
	n^{1/2}(\hat\tau_{\wald,M}-\tau_M)/ \sigma_{\wald,M} \xrightarrow{d}N(0, 1 )$. Furthermore, $\hat\sigma^2_{\wald,M}$ converges in probability to
	$$
	\lim_{n\rightarrow\infty}\frac{n}{\tau_H^2}\Big\{\frac{S_{B(1)}^2}{n_1}+\frac{S_{B(0)}^2}{n_0}\Big\} \geq  \lim_{n \rightarrow \infty} \sigma^2_{\wald,M};
	$$
	
	(2) Under Condition~\ref{cond MCATE} (1)--(4), $\hat\tau_{\ils,M}-\tau_M\xrightarrow{p}0$ and $
	n^{1/2}(\hat\tau_{\ils,M}-\tau_M)/ \sigma_{\ils,M} \xrightarrow{d}N(0,1)$.  Furthermore, $\hat\sigma^2_{\ils,M}$ converges in probability to
	$$
	\lim_{n\rightarrow\infty}\frac{n}{\tau_H^2}\Big\{\frac{S_{B_{\ols}(1)}^2}{n_1}+\frac{S_{B_{\ols}(0)}^2}{n_0}\Big\} \geq \lim_{n\rightarrow\infty} \sigma^2_{\ils,M};
	$$
	
	(3) Under Condition~\ref{cond MCATE} (1), (2), (5), (6),  $\hat\tau_{\logit,M}-\tau_M\xrightarrow{p}0$ and $
	n^{1/2}(\hat\tau_{\logit,M}-\tau_M)/ \sigma_{\logit,M} \xrightarrow{d}N(0,1)$. Furthermore, $\hat\sigma^2_{\logit,M}$ converges in probability to
	$$
	\lim_{n\rightarrow\infty}\frac{n}{\tau_H^2}\Big\{\frac{S_{B_{\logit}(1)}^2}{n_1}+\frac{S_{B_{\logit}(0)}^2}{n_0}\Big\} \geq 	\lim_{n\rightarrow\infty} \sigma^2_{\logit,M};
	$$
	
	(4) Under Condition~\ref{cond MCATE} (1), (2), (5), (7), $\hat\tau_{\cal,M}-\tau_M\xrightarrow{p}0$ and $
	n^{1/2}(\hat\tau_{\cal,M}-\tau_M) / \sigma_{\cal,M} \xrightarrow{d}N(0,1)$. Furthermore, $\hat\sigma^2_{\cal,M}$ converges in probability to
	$$
	\lim_{n\rightarrow\infty}\frac{n}{\tau_H^2}\Big\{\frac{S_{B_{\cal}(1)}^2}{n_1}+\frac{S_{B_{\cal}(0)}^2}{n_0}\Big\} \geq \lim_{n\rightarrow\infty} \sigma^2_{\cal,M}.
	$$
\end{theorem}

Theorem \ref{thm::MCATE} provides normal approximations for the distributions of the proposed MCATE estimators. The variance estimators are generally conservative and they are consistent if and only if the unit-level treatment effects on the transformed potential outcomes are constant.

\begin{theorem}\label{thm::MCATE var}
	(1)	The difference between the asymptotic variances of $n^{1/2}\hat\tau_{\ils,M}$ and $n^{1/2}\hat\tau_{\wald,M}$ is 
	$$
	-\lim_{n\rightarrow\infty}\frac{1}{\tau_H^2 p_{1} p_{0}} \{p_{0} \delta_M(1)+p_{1} \delta_M(0)\}^{\T} S_{\X}^2\{p_{0} \delta_M(1)+p_{1} \delta_M(0)\}\leq 0,
	$$
	and the difference between the variance estimators $\hat\sigma^2_{\ils,M}$ and $\hat\sigma^2_{\wald,M}$ converges in probability to
	$$
	-\lim_{n\rightarrow\infty}\frac{1}{\tau_H^2}\Big\{\frac{\delta_M(1)^{\T} S_{\X}^2 \delta_M(1)}{p_{1}}+\frac{\delta_M(0)^{\T}  S_{\X}^2  \delta_M(0)}{p_{0}}\Big\}\leq 0,
	$$
	where $\delta_M(1)=\beta_G(1)-\tau\beta_H(1)$ and  $\delta_M(0)=\beta_G(0)-\tau\beta_H(0)$;
	
	(2) The difference between the asymptotic variances of  $n^{1/2}\hat\tau_{\cal,M}$ and $n^{1/2}\hat\tau_{\wald,M}$ is
	$$
	-\lim_{n\rightarrow\infty}\frac{1}{\tau_H^2p_{1} p_{0}}\{p_{0} \phi_M(1)+p_{1} \phi_M(0)\}^{\T} S_{\V }^2\{p_{0} \phi_M
	(1)+p_{1} \phi_M(0)\}\leq 0,
	$$
	and the difference between the  variance estimators  $\hat\sigma_{\cal,M}^2$ and $\hat\sigma^2_{\wald,M}$ converges in probability to
	$$
	-\lim_{n\rightarrow\infty}\frac{1}{\tau_H^2}\Big\{\frac{\phi_M(1)^{\T} S_{\V }^2 \phi_M(1)}{p_{1}}+\frac{\phi_M(0)^{\T}  S_{\V}^2  \phi_M(0)}{p_{0}}\Big\}\leq 0,
	$$
	where $\phi_M(1)=\gamma_G(1)-\tau\gamma_H(1)$ and  $\phi_M(0)=\gamma_G(0)-\tau\gamma_H(0)$;
	
	(3) The asymptotic variance of $\hat\tau_{\cal,M}$ and its variance estimator are less than or equal to those of $\hat\tau_{\logit,M}$.
\end{theorem}

Theorem~\ref{thm::MCATE var} shows that, $\hat\tau_{\ils,M}$ and $\hat\tau_{\cal,M}$ generally improve the efficiency compared with $\hat\tau_{\wald,M}$, and $\hat\tau_{\cal,M}$ generally improves the efficiency compared with $\hat\tau_{\logit,M}$. The relative efficiency of $\hat\tau_{\ils,M}$ and $\hat\tau_{\cal,M}$ depends on which models,  linear or logistic, fit the data better.

\section{Proof of main results}

\subsection{Proof of Theorem \ref{thm::unadj}}

\begin{proof}
	
	Our proof relies on the finite-population central limit theorem (CLT) developed by \citet{LiDing2017}; see the following Lemma~\ref{lem::CLT}.
	
	In completely randomized experiments, let $R_i(z)$ be the potential outcome with observed value $R_i^\obs$, and $\tau_R$ be the ITT effect of $Z$ on $R$. The difference-in-means estimator for $\tau_R$ is $\hat\tau_R=\bar R_1^{\obs}-\bar R_0^{\obs}$, and the variance of $\hat\tau_R$ is $\var( \hat\tau_R ) = S_{R(1)}^2/n_1+S_{R(0)}^2/n_0-S_{R(1)-R(0)}^2/n$. Let $\sigma^2_R = n \var( \hat\tau_R )$, and $s_{R(z)}^2$ and $s_{\X R(z)}^2$ be the sample variance and covariance under treatment arm $z$, $z=0,1$. Let $Q_i(z)$ be another potential outcome. 
	
	
	\begin{lemma}\label{lem::CLT}
		
		As $n\rightarrow\infty$, for $z=0,1$, if (i) $n_z/n$ has positive limiting value between 0 and 1, i.e., $n_1/n\rightarrow p_1\in(0,1)$ and $n_0/n\rightarrow p_0\in(0,1)$; (ii) $\max_{i\in \{1,...,n\}} \{R_i(z)-\bar R(z)\}^2/n\rightarrow 0$; (iii) $S_{R(z)}^2$ and $S_{R(1)-R(0)}^2$ have limiting values and the limit of $\sigma^2_R$ is positive, then $\bar R_z^{\obs} - \bar{R}(z)\xrightarrow{p}0$, $s_{R(z)}^2-S_{R(z)}^2 \xrightarrow{p}0$,  $\hat\tau_R-\tau_R\xrightarrow{p}0$, and $n^{1/2}(\hat\tau_R-\tau_R) / \sigma_R \xrightarrow{d}N(0,1)$. Furthermore, if $\max_{i\in \{1,...,n\}} \{Q_i(z)-\bar Q(z)\}^2/n\rightarrow 0$, then $s_{ R(z) Q(z) }^2-S_{ R(z) Q(z)}^2 \xrightarrow{p}0$.
		
	\end{lemma}

	
	Under Conditions~\ref{Identification assumptions} and \ref{cond unadj} and by Lemma~\ref{lem::CLT}, we have $\tau_D = n_c/n > C_0 > 0$, $\hat\tau_Y - \tau_Y \xrightarrow{p} 0 $, and $\hat\tau_D - \tau_D \xrightarrow{p} 0 $. Thus, using Slutsky's theorem, we have $\hat\tau_{\wald}- \tau \xrightarrow{p} 0 $ and $\hat\tau_{\wald}-\tau=(\hat\tau_{Y}-\tau\hat\tau_{D})/\hat\tau_D$ has the same asymptotic distribution as $(\hat\tau_{Y}-\tau\hat\tau_{D})/\tau_D=\hat\tau_A/\tau_D$, where $\hat\tau_A$ is the difference-in-means estimator for the ITT effect of $Z$ on the transformed potential outcomes $A$: $A_i(z)=Y_i(z)-\tau D_i(z)$, $z=0,1$.
	
	It is easy to check that $\tau_A = n^{-1}\sum_{i=1}^n \{ A_i(1) - A_i(0) \} = \tau_Y - \tau \tau_D = 0$. Moreover, $\tau = \tau_Y / \tau_D \leq 1/ \tau_D < 1 / C_0 < \infty$. As $Y_i(z)$ and $D_i(z)$ are binary, $A_i(z)$ is bounded. Thus,  the regularity condition on the maximum squared distance $\max_{i\in \{1,...,n\}} \{A_i(z)-\bar A(z)\}^2/n\rightarrow 0$ holds. Under Condition~\ref{cond unadj}, the finite-population variances of $A_i(1)$, $A_i(0)$ and $A_i(1)-A_i(0)$,
	$$
	S_{A(1)}^2=S_{Y(1)}^2+\tau^2S_{D(1)}^2-2\tau S_{Y(1)D(1)},
	$$
	$$
	S_{A(0)}^2=S_{Y(0)}^2+\tau^2S_{D(0)}^2-2\tau S_{Y(0)D(0)},
	$$
	$$
	S_{A(1)-A(0)}^2=S_{Y(1)-Y(0)}^2+\tau^2S_{D(1)-D(0)}^2-2\tau S_{\{Y(1)-Y(0)\}\{D(1)-D(0)\}},
	$$
	have limiting values. Applying the finite-population CLT  (Lemma~\ref{lem::CLT}) to $A_i(z)$, we have $n^{1/2}(\hat\tau_A-\tau_A)/\sigma_A\xrightarrow{d}N(0,1)$, where
	$$
	\sigma^2_A=\frac{S_{A(1)}^2}{n_1/n}+\frac{S_{A(0)}^2}{n_0/n}-S_{A(1)-A(0)}^2.
	$$
	Therefore, $n^{1/2}(\hat\tau_{\wald}-\tau)/\sigma_{\wald}\xrightarrow{d}N(0,1)$, where $\sigma^2_{\wald}=\sigma^2_A/\tau_D^2$. 
	
	By Lemma~\ref{lem::CLT}, we  also have  $s_{Y(z)}^2-S_{Y(z)}^2\xrightarrow{p}0$, $s_{D(z)}^2-S_{D(z)}^2\xrightarrow{p}0$, and  $s_{Y(z)D(z)}-S_{Y(z)D(z)}\xrightarrow{p}0$. Since
	$$
	s_{A(z)}^2 = \frac{1}{n_z - 1} \sum_{i: Z_i = z} \Big\{ \hat A_i(z) - \frac{1}{n_z} \sum_{j: Z_j = z}  \hat A_{j}(z) \Big\}^2 = s_{Y(z)}^2 + \hat\tau_{\wald}^2 s_{D(z)}^2 - 2 \hat\tau_{\wald} s_{Y(z)D(z)}, 
	$$
	$$
	S_{A(z)}^2 = \frac{1}{n - 1} \sum_{i=1}^{n} \Big\{ A_i(z) - \bar{A}(z) \Big\}^2 = S_{Y(z)}^2 + \tau^2 S_{D(z)}^2 - 2 \tau S_{Y(z)D(z)}, 
	$$
	then, $s_{A(z)}^2 - S_{A(z)}^2 \xrightarrow{p}0$. Therefore,
	$$
	\hat \sigma^2_{\wald} \xrightarrow{p}\lim_{n\rightarrow\infty}\frac{n}{\tau_D^2}\Big\{\frac{S_{A(1)}^2}{n_1}+\frac{S_{A(0)}^2}{n_0}\Big \} \geq \lim_{n\rightarrow\infty} \sigma^2_{\wald}.
	$$
	
\end{proof}

\subsection{Proof of Theorem \ref{thm::ols}}

\begin{proof}
	
	First, we present a lemma on the asymptotic properties of the OLS-adjusted estimator $\hat \tau_Y$ (or $\hat \tau_D$).
	Let us denote $\hat\tau_{\ols,R}$ as the OLS-adjusted estimator for $\tau_R$,  
	$$
	\hat\tau_{\ols,R}=\Big\{\bar R_1^{\obs}-(\bar{\X}_1^{\obs}-\bar{\X})^{\T}\hat\beta_R(1) \Big\}- \Big\{\bar R_0^{\obs}-(\bar{\X}_0^{\obs}-\bar{\X})^{\T}\hat\beta_R(0)\Big\},
	$$
	where 
	$$
	\hat{\beta}_{R}(z)=\underset{\beta\in\mathbb{R}^{P}}{\arg \min } \sum_{i: Z_i=z} \{R_{i}^{\obs}-\bar{R}_z^{\obs}-(\X_i-\bar{\X}_z^{\obs})^{\T} \beta\}^{2}, \quad z=0,1.
	$$
	We define the population regression (projection) coefficient vector $\beta_R(z)=(S_\X^2)^{-1}S_{\X R(z)}$ and decompose the potential outcomes as $R_i(z)=\bar R(1)+(\X_i-\bar{\X})^{\T}\beta_R(z)+\epsilon_{R,i}(z)$. Let $\sigma_{\ols,R}^2=n S_{\epsilon_{R}(1)}^2/n_1+ n S_{\epsilon_{R}(0)}^2/n_0-S_{\epsilon_{R}(1)-\epsilon_{R}(0)}^2$. 
	
	\begin{lemma}\label{lem::RA}
		
		As $n\rightarrow\infty$, for $z=0,1$, if (i) $n_z/n$ has positive limiting value between 0 and 1; (ii) $\max_{i\in \{1,...,n\}} \{R_i(z)-\bar R(z)\}^2/n\rightarrow 0$ and $\max_{i\in \{1,...,n\}}\{X_{ik}-\bar{X}_{k}\}^2/n \rightarrow 0$, $k=1,\dots,p$; (iii) $S_{R(z)}^2$, $S_{R(1)-R(0)}^2$, $S_{\X}^2$, and $S_{\X R(z)}$ have limiting values with $S_{\X}^2$ and its limit being strictly positive definite, and the limit of $\sigma_{\ols,R}^2$ is positive, then we have $ \hat{\beta}_{R}(z) - \beta_R(z) \xrightarrow{p}0 $, $\hat\tau_{\ols,R} - \tau_R \xrightarrow{p} 0$ and $n^{1/2}(\hat\tau_{\ols,R}-\tau_R)/\sigma_{\ols,R}\xrightarrow{d}N(0,1)$. Furthermore, $s_{\epsilon_{R}(z)}^2-S_{\epsilon_{R}(z)}^2\xrightarrow{p}0$, where $s_{\epsilon_{R}(z)}^2$ is the sample variance of $R_i(z)-(\X_i-\bar{\X})^{\T}\hat\beta_R(z)$ under treatment arm $z$.
		
	\end{lemma}
	

	The proof of Lemma~\ref{lem::RA} can be found in \citet{LiDing2017}, so we omit. Lemma~\ref{lem::RA} implies that $\hat\tau_{\ols,R} - \tau_R \xrightarrow{p}0$ for $R = Y, D$. As $\tau_D > C_0 > 0$, we have
	$\hat\tau_{\ils} - \tau = \hat\tau_{\ols,Y} / \hat\tau_{\ols,D} - \tau_Y / \tau_D \xrightarrow{p} 0$.

	Next, we prove the asymptotic normality of $\hat\tau_{\ils}$. 
	By definition, $A_{\ols,i}(z) = Y_i(z)-\tau D_i(z)-(\X_i-\bar{\X})^{\T}\{\beta_Y(z)-\tau \beta_D(z)\}$, where
	$$
	\beta_Y(z)=(S_\X^2)^{-1}S_{\X Y(z)}, \quad \beta_D(z)=(S_\X^2)^{-1}S_{\X D(z)}, \quad z=0,1.
	$$
	Clearly,  $\beta_Y(z)-\tau\beta_D(z)$ is the slope of the population regression (or projection) of $Y_i(z)-\tau D_i(z)$ on $\X_i$. 
	Let $\hat\tau_{\ols,A}$ be the OLS-adjusted estimator for $\tau_{A}$ ($A_i(z) = Y_i(z) - \tau D_i(z)$); that is,
	\begin{eqnarray}
	\hat\tau_{\ols,A} &=& \Big[ ( \bar Y_1^{\obs}-\tau D_1^{\obs} ) - (\bar{\X}_1^{\obs}-\bar{\X})^{\T} \big\{\hat\beta_Y(1)-\tau \hat\beta_D(1)\big\} \Big] \nonumber \\
	&& - \Big[ (\bar Y_0^{\obs}-\tau D_0^{\obs})-(\bar{\X}_0^{\obs}-\bar{\X})^{\T}\big\{\hat\beta_Y(0)-\tau \hat\beta_D(0)\big\} \Big]. \nonumber
	\end{eqnarray}
	
	Under Conditions~\ref{cond unadj} and \ref{cond ols}, $\beta_R(z)$, $R=Y,D$, $z=0,1$, and $\tau$ have finite limiting values. Thus, $\max_{i\in \{1,...,n\}} \{A_{\ols,i}(z)-\bar A_{\ols}(z)\}^2/n\rightarrow 0$. Moreover, Conditions~\ref{cond unadj} and \ref{cond ols}  imply that
	\begin{eqnarray}
	S_{A_{\ols}(z)}^2 &=& S_{{A}(z)}^2 + \big\{ \beta_Y(z) - \tau \beta_D(z) \big\}^\T S_\X^2 \big\{ \beta_Y(z) - \tau \beta_D(z) \big\} \nonumber\\
	&& - 2 \big\{ \beta_Y(z) - \tau \beta_D(z) \big\}^\T \big\{ S_{ \X Y(z) } - \tau S_{ \X D(z) } \big\}, \quad z=0,1, \nonumber
	\end{eqnarray}
	have finite limits. Similarly, $S_{A_{\ols}(1)- A_{\ols}(0)}^2$ and $S_{\X A_{\ols}(z)}$ have finite limits. As the limit of
	$$
	\sigma^2_{\ols,A}=\frac{S_{A_{\ols}(1)}^2}{n_1/n}+\frac{S_{A_{\ols}(0)}^2}{n_0/n}-S_{A_{\ols}(1)-A_{\ols}(0)}^2 = \tau_D^2 \sigma^2_{\ils}
	$$
	is positive, applying Lemma~\ref{lem::RA} to $\hat\tau_{\ols,A}$, we have $n^{1/2}(\hat\tau_{\ols,A}-\tau_{A})/\sigma_{\ols,A}\xrightarrow{d}N(0,1)$.
	Moreover, $s_{A_{\ols}(z)}^2 - S_{A_{\ols}(z)}^2 \xrightarrow{p}0 $, $z=0,1$.

	As $\hat\tau_{\ils}-\tau=\hat\tau_{\ols,A}/\hat\tau_{\ols,D}$ and $\hat\tau_{\ols,D} - \tau_D \xrightarrow{p} 0$, by Slutsky's theorem, we have  $n^{1/2}(\hat\tau_{\ils}-\tau)/\sigma_{\ils}\xrightarrow{d}N(0,1)$. Moreover, 
	$$
	\hat \sigma^2_{\ils} = \frac{n}{\hat\tau_{\ols,D}^2}  \Big\{\frac{s_{A_{\ols}(1)}^2}{n_1}+\frac{s_{A_{\ols}(0)}^2}{n_0}\Big\}  \xrightarrow{p}\lim_{n\rightarrow\infty}\frac{n}{\tau_D^2}\Big\{\frac{S_{A_{\ols}(1)}^2}{n_1}+\frac{S_{A_{\ols}(0)}^2}{n_0}\Big\}\geq \lim_{n\rightarrow\infty}\sigma_{\ils}^2.
	$$

	
\end{proof}

\subsection{Proof of Theorem \ref{thm::ols var}}

\begin{proof}
	
	According to Theorems~\ref{thm::unadj} and \ref{thm::ols}, the asymptotic variances of $n^{1/2}\hat\tau_{\wald}$ and $n^{1/2}\hat\tau_{\ils}$ are, respectively,
	$$
	\lim_{n\rightarrow\infty}\frac{n}{\tau_D^2}\Big\{\frac{S_{A(1)}^2}{n_1}+\frac{S_{A(0)}^2}{n_0}- \frac{S_{A(1)-A(0)}^2}{n}\Big\},
	$$
	$$
	\lim_{n\rightarrow\infty}\frac{n}{\tau_D^2}\Big\{\frac{S_{A_{\ols}(1)}^2}{n_1}+\frac{S_{A_{\ols}(0)}^2}{n_0}- \frac{S_{A_{\ols}(1)-A_{\ols}(0)}^2}{n}\Big\}.
	$$
	Note that, for $z=0,1$,
	$$
	\begin{aligned}
	A_{\ols,i}(z)&=Y_i(z)-\tau D_i(z)-(\X_i-\bar {\X})^{\T}\{\beta_Y(z)-\tau\beta_D(z)\}\\
	&=A_i(z)-(\X_i-\bar {\X})^{\T}\{\beta_Y(z)-\tau\beta_D(z)\},
	\end{aligned}
	$$
	where $A_i(z)=Y_i(z)-\tau D_i(z)$. Thus,
	\begin{eqnarray}
	S_{A_{\ols}(1)}^2 &= & S_{A(1)}^2+\{\beta_Y(1)-\tau\beta_D(1)\}^{\T}S_{\X}^2\{\beta_Y(1)-\tau\beta_D(1)\} \nonumber \\
	&& - 2 \big\{ \beta_Y(1) - \tau \beta_D(1) \big\}^\T \big\{ S_{ \X Y(1) } - \tau S_{ \X D(1) } \big\} \nonumber \\
	& = & S_{A(1)}^2 - \{\beta_Y(1)-\tau\beta_D(1)\}^{\T}S_{\X}^2\{\beta_Y(1)-\tau\beta_D(1)\}. \nonumber
	\end{eqnarray}
	Similarly,
	$$
	S_{A_{\ols}(0)}^2=S_{A(0)}^2-\{\beta_Y(0)-\tau\beta_D(0)\}^{\T}S_{\X}^2\{\beta_Y(0)-\tau\beta_D(0)\}.
	$$
	Let $\delta(1)=\beta_Y(1)-\tau\beta_D(1)$ and  $\delta(0)=\beta_Y(0)-\tau\beta_D(0)$, then 
	$$
	S_{A_{\ols}(1)}^2=S_{A(1)}^2-\delta(1)^{\T}S_{\X}^2\delta(1),
	\quad 
	S_{A_{\ols}(0)}^2=S_{A(0)}^2-\delta(0)^{\T}S_{\X}^2\delta(0).
	$$
	As
	$
	A_{\ols,i}(1)-A_{\ols,i}(0)=A_i(1)-A_i(0)-(\X_i-\bar {\X})^{\T}\{\delta(1)-\delta(0)\},
	$
	we have
	$$
	\begin{aligned}
	S_{A_{\ols}(1)-A_{\ols}(0)}^2=&S_{A(1)-A(0)}^2+\{\delta(1)-\delta(0)\}^{\T}S_{\X}^2\{\delta(1)-\delta(0)\}\\&-\frac{2}{n-1}\sum_{i=1}^{n}\{A_i(1)-A_i(0)\}(\X_i-\bar {\X})^{\T}\{\delta(1)-\delta(0)\}\\
	=&S_{A(1)-A(0)}^2-\{\delta(1)-\delta(0)\}^{\T}S_{\X}^2\{\delta(1)-\delta(0)\}.
	\end{aligned}
	$$
	Therefore, the difference between the asymptotic variances of $n^{1/2}\hat\tau_{\ils}$ and $n^{1/2}\hat\tau_{\wald}$ is the limit of $1/\tau_D^2$ times
	$$
	\begin{aligned}
	&-\frac{\delta(1)^{\T} S_{\X}^2 \delta(1)}{p_{1}}-\frac{\delta(0)^{\T}  S_{\X}^2  \delta(0)}{p_{0}}+\{\delta(1)-\delta(0)\}^{\T}  S_{\X}^2 \{\delta(1)-\delta(0)\} \\
	=&-\frac{1}{p_{1} p_{0}}\big[p_{0} \delta(1)^{\T} S_{\X}^2 \delta(1)+p_{1} \delta(0)^{\T} S_{\X}^2\delta(0)-p_{1} p_{0}\{\delta(1)-\delta(0)\}^{\T} S_{\X}^2\{\delta(1)-\delta(0)\} \big]\\
	=&-\frac{1}{p_{1} p_{0}} \big[ p_{0}^{2} \delta(1)^{\T} S_{\X}^2\delta(1)+p_{1}^{2} \delta(0)^{\T} S_{\X}^2\delta(0)+2 p_{1} p_{0} \delta(1)^{\T} S_{\X}^2 \delta(0) \big] \\
	=&-\frac{1}{p_{1} p_{0}}\{p_{0} \delta(1)+p_{1} \delta(0)\}^{\T} S_{\X}^2\{p_{0} \delta(1)+p_{1} \delta(0)\} \\
	\leq& 0.
	\end{aligned}
	$$
	Furthermore, since
	$$
	\hat \sigma^2_{\wald} \xrightarrow{p}\lim_{n\rightarrow\infty}\frac{n}{\tau_D^2}\Big\{\frac{S_{A(1)}^2}{n_1}+\frac{S_{A(0)}^2}{n_0}\Big \},
	\quad
	\hat \sigma^2_{\ils} \xrightarrow{p}\lim_{n\rightarrow\infty}\frac{n}{\tau_D^2}\Big\{\frac{S_{A_{\ols}(1)}^2}{n_1}+\frac{S_{A_{\ols}(0)}^2}{n_0}\Big \},
	$$
	and $n_z / n \rightarrow p_z$, $z=0,1$,
	then, the difference between the variance estimators $\hat\sigma^2_{\ils}$ and $\hat\sigma^2_{\wald}$ converges in probability to
	$$
	-\lim_{n\rightarrow\infty}\frac{1}{\tau_D^2}\Big\{\frac{\delta(1)^{\T} S_{\X}^2 \delta(1)}{p_{1}}+\frac{\delta(0)^{\T}  S_{\X}^2  \delta(0)}{p_{0}}\Big\}\leq 0.
	$$
	
\end{proof}

\subsection{Proof of Theorem \ref{thm::logit}}

\begin{proof}
	
	First, we introduce a useful result on the asymptotic linear expansion of the imputed potential outcomes, obtained by \citet{Guo2021}.

	\begin{lemma}\label{lem::ALE}
		
		Under Conditions~\ref{Identification assumptions}, \ref{cond unadj} and \ref{cond logit}, we have
		$$
		\left\|\hat{\mu}_{R}(z)-\mu_{R}(z)\right\|_{n}^2 = \frac{1}{n} \sum_{i=1}^{n}  \big\{ \hat{\mu}_{R,i}(z)-\mu_{R,i}(z) \big\}^2 \xrightarrow{p}0,
		$$
		$$
		\frac{1}{n}\sum_{i=1}^n \{\hat R_i(z)-R_i(z)\}=\frac{1}{n_z}\sum_{i:Z_i=z} \eta_{R,i}(z)+o_p(n^{-1/2}), \quad R = Y,D, \quad z = 0, 1.
		$$
		
	\end{lemma}
	
	Lemma~\ref{lem::ALE} is a direct result of Theorems 2 and 3 of \citet{Guo2021} applied to $Y_i(z)$ and $D_i(z)$, respectively, so we omit the proof. Next, we prove the theorem.

	
	According to Lemma~\ref{lem::ALE}, we have
	$$
	\hat\tau_{\logit,Y} - \tau_Y = \frac{1}{n_1} \sum_{i:Z_i = 1} \eta_{Y,i}(1) - \frac{1}{n_0} \sum_{i:Z_i = 0} \eta_{Y,i}(0)  +o_p(n^{-1/2}).
	$$
	According to the definite of $\eta_{Y,i}(z)$, we have $\bar{\eta}_Y(z) = 0$ (by first-order condition of the optimization problem). Then, by Lemma~\ref{lem::CLT}, under Condition~\ref{cond logit}, 
	$$
	\frac{1}{n_1} \sum_{i:Z_i = 1} \eta_{Y,i}(1) - \frac{1}{n_0} \sum_{i:Z_i = 0} \eta_{Y,i}(0) = o_p(1).
	$$
	Therefore, $\hat\tau_{\logit,Y} - \tau_Y \xrightarrow{p} 0 $. Similarly, $\hat\tau_{\logit,D} - \tau_D \xrightarrow{p} 0$. Then, $\hat\tau_{\logit} - \tau =\hat\tau_{\logit,Y}/\hat\tau_{\logit,D} - \tau_Y / \tau_D \xrightarrow{p} 0$. 
	
	Recall that
	$$
	A_{\logit,i}(z)=Y_i(z)-\tau D_i(z)-\{\mu_{Y,i}(z)-\tau \mu_{D,i}(z)\}=\eta_{Y,i}(z)-\tau \eta_{D,i}(z).
	$$
	Then,
	$$
	\begin{aligned}
	&\hat\tau_{\logit,Y}-\tau\hat\tau_{\logit,D}\\
	=&\frac{1}{n}\sum_{i=1}^n\hat Y_i(1)-\frac{1}{n}\sum_{i=1}^n\hat Y_i(0)-\tau\Big\{\frac{1}{n}\sum_{i=1}^n\hat D_i(1)-\frac{1}{n}\sum_{i=1}^n\hat D_i(0)\Big\}\\
	=&\frac{1}{n_1}\sum_{i:Z_i=1} \eta_{Y,i}(1)-\frac{1}{n_0}\sum_{i:Z_i=0} \eta_{Y,i}(0)-\tau\Big\{\frac{1}{n_1}\sum_{i:Z_i=1} \eta_{D,i}(1)-\frac{1}{n_0}\sum_{i:Z_i=0} \eta_{D,i}(0)\Big\}+o_p(n^{-1/2})\\
	=&\frac{1}{n_1}\sum_{i:Z_i=1}A_{\logit,i}(1)-\frac{1}{n_0}\sum_{i:Z_i=0} A_{\logit,i}(0)+o_p(n^{-1/2}) \\
	= & \hat\tau_{A_{\logit}} +o_p(n^{-1/2}),
	\end{aligned} 
	$$
	where $\hat\tau_{A_{\logit}} =\sum_{i:Z_i=1}A_{\logit,i}(1)/n_1-\sum_{i:Z_i=0} A_{\logit,i}(0)/n_0$ is the difference-in-means estimator for the ITT effect of $Z$ on $A_\logit$ with $\tau_{A_{\logit}} = 0$. First, since $Y_i(z)$ and $D_i(z)$ are binary and the predicted probabilities $\mu_{Y,i}(z)$ and $\mu_{D,i}(z)$ are bounded, then, $A_{\logit,i}(z)$ is bounded. Thus,  $\max_{i\in \{1,...,n\}} \{A_{\logit,i}(z)-\bar A_{\logit}(z)\}^2/n\rightarrow 0$. Second, under Condition \ref{cond logit}, the finite-population variances, $S_{A_{\logit}(1)}^2$, $S_{A_{\logit}(0)}^2$, and $S_{A_{\logit}(1)-A_{\logit}(0)}^2$, tend to finite limits.
	Then, applying the finite-population CLT (Lemma~\ref{lem::CLT}) to the transformed potential outcomes $A_{\logit,i}(z)$, we have $n^{1/2} \hat\tau_{A_{\logit}} /\sigma_{A_{\logit}}\xrightarrow{d}N(0,1)$, where
	$$
	\sigma_{A_{\logit}}^2=\frac{S_{A_{\logit}(1)}^2}{n_1/n}+\frac{S_{A_{\logit}(0)}^2}{n_0/n}-S_{A_{\logit}(1)-A_{\logit}(0)}^2.
	$$
	Since $\hat\tau_{\logit} - \tau = ( \hat\tau_{\logit,Y}-\tau \hat\tau_{\logit,D} ) / \hat\tau_{\logit,D} = \hat\tau_{A_{\logit}} / \hat\tau_{\logit,D} +  o_p(n^{-1/2}) $  follows the same asymptotic distribution as $\hat\tau_{A_{\logit}} / \tau_D$, then, $n^{1/2}(\hat\tau_{\logit}-\tau)/\sigma_{\logit}\xrightarrow{d}N(0,1)$, where $\sigma^2_{\logit}=\sigma_{A_{\logit}}^2/\tau_D^2.$ 
	
	
	
	In addition, by Lemma~\ref{lem::ALE},  $\left\|\hat{\mu}_{Y}(z)-\mu_{Y}(z)\right\|_{n}^2 \xrightarrow{p}0,$ then,
	$$
	\begin{aligned}
	s_{\eta_{Y}(z)}^2=&\frac{1}{n_z-p-1}\sum_{i: Z_i=z} \left\{Y_{i}(z)-\hat\mu_{Y,i}(z)\right\}^{2}\\
	=&\frac{1}{n_z-p-1}\sum_{i: Z_i=z}\Big[\{Y_{i}(z)-\mu_{Y,i}(z)\}+ \{\mu_{Y,i}(z)-\hat\mu_{Y,i}(z)\}\Big]^2\\
	\xrightarrow{p}&\lim_{n\rightarrow\infty} S_{\eta_Y(z)}^2.
	\end{aligned}
	$$
	Similarly,  $s_{\eta_D(z)}^2-S_{\eta_D(z)}^2\xrightarrow{p}0$, and  $s_{\eta_Y(z)\eta_D(z)}-S_{\eta_Y(z)\eta_D(z)}\xrightarrow{p}0$. Thus,
	\begin{eqnarray}
	s_{A_{\logit}(z)}^2 & = & s_{\eta_{Y}(z)}^2 + \hat\tau_{\logit}^2 s_{\eta_D(z)}^2 - 2 \hat\tau_{\logit} s_{\eta_Y(z)\eta_D(z)} \nonumber \\
	&\xrightarrow{p} & \lim_{n\rightarrow\infty} \big\{ S_{\eta_{Y}(z)}^2 + \tau^2 S_{\eta_D(z)}^2 - 2 \tau S_{\eta_Y(z)\eta_D(z)} \big\} = \lim_{n\rightarrow\infty} S_{A_{\logit}(z)}^2. \nonumber
	\end{eqnarray}
	Therefore,
	$$
	\hat \sigma^2_{\logit} \xrightarrow{p}\lim_{n\rightarrow\infty}\frac{n}{\tau_D^2}\Big\{\frac{S_{A_{\logit}(1)}^2}{n_1}+\frac{S_{A_{\logit}(0)}^2}{n_0}\Big \} \geq \lim_{n\rightarrow\infty}\sigma^2_{\logit}.
	$$
	
\end{proof}

\subsection{Proof of Theorem \ref{thm::cal}}

\begin{proof}

	For $z = 0, 1$, we define $\mu_{\cal,Y,i}(z)=\bar Y(z)+(\W_{i}-\bar{\W})^{\T}\gamma_Y(z)$ as the prediction equation from the  population-level linear regression of $Y_{i}(z)$ on $\W_{i}$ with an intercept, and $\hat{\mu}_{\cal,Y,i}(z)=\bar Y_z^{\obs}+(\W_{i}^{\obs}-\bar{\W}_z^{\obs})^{\T}\hat\gamma_Y(z)$ as the prediction equation from the sample-level linear regression of $Y_{i}^{\obs}$ on $\W_{i}$ with an intercept under treatment arm $z$.
	Similarly, we define $\mu_{\cal,D,i}(z)$  and $\hat{\mu}_{\cal,D,i}(z)$. We use $\hat \mu_{\cal,Y,i}(z)$ and $\hat \mu_{\cal,D,i}(z)$ to  impute the unobserved potential outcomes, and let 
	$$
	\hat{Y}_{\cal,i}(1)=\Bigg\{\begin{array}{ll}Y_{i}(1)  &\ Z_{i}=1 \\ \hat{\mu}_{\cal,Y,i}(1) &\ Z_{i}=0
	\end{array} ,
	\quad 
	\hat{Y}_{\cal,i}(0)=\Bigg\{\begin{array}{ll}\hat{\mu}_{\cal,Y,i}(0) &\ Z_{i}=1 \\ Y_{i}(0)  &\ Z_{i}=0
	\end{array},
	$$
	$$
	\hat{D}_{\cal,i}(1)=\Bigg\{\begin{array}{ll}D_{i}(1)  &Z_{i}=1 \\ \hat{\mu}_{\cal,D,i}(1) &Z_{i}=0
	\end{array} ,
	\quad 
	\hat{D}_{\cal,i}(0)=\Bigg\{\begin{array}{ll}\hat{\mu}_{\cal,D,i}(0) &Z_{i}=1 \\ D_{i}(0)  &Z_{i}=0
	\end{array}.
	$$
	Then, the calibrated Oaxaca--Blinder estimators for $\tau_Y$ and $\tau_D$ have the following form
	$\hat\tau_{\cal,Y}=\sum_{i=1}^n\{\hat{Y}_{\cal,i}(1)-\hat{Y}_{\cal,i}(0)\}/n$ and $\hat\tau_{\cal,D}=\sum_{i=1}^n\{\hat{D}_{\cal,i}(1)-\hat{D}_{\cal,i}(0)\}/n$.

	Applying an intermediate result in the proof of Theorem 1 of \cite{Cohen2020} to the calibrated Oaxaca--Blinder estimators $\hat\tau_{\cal,Y}$ and $\hat\tau_{\cal,Y}$ separately, we have the following lemma:
	
	\begin{lemma}\label{lem::CAL}
		
		Under Conditions~\ref{Identification assumptions}, \ref{cond unadj}, \ref{cond logit}, and \ref{cond cal}, we have
		$$
		\left\|\hat{\mu}_{\cal,R}(z)-\mu_{\cal,R}(z)\right\|_{n}^2 = \frac{1}{n} \sum_{i=1}^{n}  \big\{ \hat{\mu}_{\cal,R,i}(z)-\mu_{\cal,R,i}(z) \big\}^2 \xrightarrow{p}0,
		$$
		$$
		\frac{1}{n}\sum_{i=1}^n \{\hat R_{\cal, i}(z)-R_{\cal,i}(z)\}=\frac{1}{n_z}\sum_{i:Z_i=z} \xi_{R,i}(1)+o_p(n^{-1/2}), \quad R = Y,D, \quad z = 0, 1,
		$$
		where $\xi_{Y,i}(z)=Y_{i}(z)-\mu_{\cal,Y,i}(z)$ and $\xi_{D,i}(z)=D_{i}(z)-\mu_{\cal,D,i}(z)$.
		
	\end{lemma}

	
	Similar to the proof of Theorem~\ref{thm::logit}, by Lemmas~\ref{lem::CLT} and \ref{lem::CAL}, we have 
	$\hat\tau_{\cal,Y} - \tau_Y \xrightarrow{p} 0$ and $\hat\tau_{\cal,D} - \tau_D \xrightarrow{p} 0$. Thus, $\hat\tau_{\cal} - \tau =\hat\tau_{\cal,Y}/\hat\tau_{\cal,D} - \tau_Y / \tau_D \xrightarrow{p} 0$. 
	
	Next, we prove the asymptotic normality. Since
	\begin{eqnarray}
	A_{\cal,i}(z) & =& Y_i(z)-\tau D_i(z)-(\W_i-\bar{\W})^{\T}\{\gamma_Y(z)-\tau \gamma_D(z) \} \nonumber \\
	&=&\bar Y(z)-\tau\bar D(z)+\xi_{Y,i}(z)-\tau \xi_{D,i}(z), \nonumber
	\end{eqnarray}
	then
	$$
	\begin{aligned}
	&\hat\tau_{\cal,Y}-\tau\hat\tau_{\cal,D}\\
	=&\frac{1}{n}\sum_{i=1}^n\hat Y_{\cal,i}(1)-\frac{1}{n}\sum_{i=1}^n\hat Y_{\cal,i}(0)-\tau\Big\{\frac{1}{n}\sum_{i=1}^n\hat D_{\cal,i}(1)-\frac{1}{n}\sum_{i=1}^n\hat D_{\cal,i}(0)\Big\}\\
	=&\frac{1}{n_1}\sum_{i:Z_i=1} \xi_{Y,i}(1)-\frac{1}{n_0}\sum_{i:Z_i=0} \xi_{Y,i}(0)-\tau\Big\{\frac{1}{n_1}\sum_{i:Z_i=1} \xi_{D,i}(1)-\frac{1}{n_0}\sum_{i:Z_i=0} \xi_{D,i}(0)\Big\}+o_p(n^{-1/2})\\
	=&\frac{1}{n_1}\sum_{i:Z_i=1}A_{\cal,i}(1)-\frac{1}{n_0}\sum_{i:Z_i=0} A_{\cal,i}(0)+o_p(n^{-1/2}) \\
	= & \hat\tau_{A_{\cal}} +o_p(n^{-1/2}),
	\end{aligned}
	$$
	where $\hat\tau_{A_{\cal}}=\sum_{i:Z_i=1}A_{\cal,i}(1)/n_1-\sum_{i:Z_i=0} A_{\cal,i}(0)/n_0$ is the difference-in-means estimator for the ITT effect of $Z$ on $A_{\cal}(z)$ with  $\tau_{A_{\cal}} = 0$. To derive the asymptotic normality of $\hat\tau_{A_{\cal}}$, we only need to check that the conditions required by the finite-population CLT hold. First, as $Y_i(z)$ and $D_i(z)$ are binary, and the covariates $\W_i$ are bounded, then $\gamma_Y(z)$ and $\gamma_D(z)$ are bounded. Thus, $A_{\cal,i}(z)$ is bounded, which implies $\max_{i\in \{1,...,n\}} \{A_{\cal,i}(z)-\bar A_{\cal}(z)\}^2/n\rightarrow 0$. Second, the finite-population variances of $A_{\cal,i}(1)$, $A_{\cal,i}(0)$, and $A_{\cal,i}(1)-A_{\cal,i}(0)$ are
	$$
	S_{A_{\cal}(1)}^2=S_{\xi_Y(1)}^2+\tau^2S_{\xi_D(1)}^2-2\tau S_{\xi_Y(1)\xi_D(1)},
	$$
	$$
	S_{A_{\cal}(0)}^2=S_{\xi_Y(0)}^2+\tau^2S_{\xi_D(0)}^2-2\tau S_{\xi_Y(0)\xi_D(0)},
	$$
	$$
	S_{A_{\cal}(1)-A_{\cal}(0)}^2=S_{\xi_Y(1)-\xi_Y(0)}^2+\tau^2S_{\xi_D(1)-\xi_D(0)}^2-2\tau S_{\{\xi_Y(1)-\xi_Y(0)\}\{\xi_D(1)-\xi_D(0)\}}.
	$$
	Under Conditions \ref{cond unadj} and \ref{cond cal}, the above variances tent to finite limits. Then, applying Lemma~\ref{lem::CLT} to $A_{\cal,i}(z)$, we have $n^{1/2} \hat\tau_{A_{\cal}}/\sigma_{A_{\cal}}\xrightarrow{d}N(0,1)$, where
	$$
	\sigma_{A_{\cal}}^2=\frac{S_{A_{\cal}(1)}^2}{n_1/n}+\frac{S_{A_{\cal}(0)}^2}{n_0/n}-S_{A_{\cal}(1)-A_{\cal}(0)}^2.
	$$
	As $\hat\tau_{\cal}-\tau = ( \hat\tau_{\cal,Y}-\tau \hat\tau_{\cal,D} ) / \hat\tau_{\cal,D} =\hat\tau_{A_{\cal}} / \hat\tau_{\cal,D} +o_p(n^{-1/2})$, by Slutsky's theorem, we have $n^{1/2}(\hat\tau_{\cal}-\tau)/\sigma_{\cal}\xrightarrow{d}N(0,1)$, where $\sigma^2_{\cal}=\sigma^2_{A_{\cal}}/\tau_D^2$.
	
	
	
	Finally, we derive the probability limit of the variance estimator. By definition and Lemma~\ref{lem::CAL},
	$$
	\begin{aligned}
	s_{\xi_{Y}(z)}^2=&\frac{1}{n_z-5}\sum_{i: Z_i=z} \left\{Y_{i}(z)-\hat\mu_{\cal,Y,i}(z)\right\}^{2}\\
	=&\frac{1}{n_z-5}\sum_{i: Z_i=z}\Big[\{Y_{i}(z)-\mu_{\cal,Y,i}(z)\}+\{\mu_{\cal,Y,i}(z)-\hat\mu_{\cal,Y,i}(z)\}\Big]^2\\
	\xrightarrow{p}&\lim_{n\rightarrow\infty} S_{\xi_Y(z)}^2.
	\end{aligned}
	$$
	Similarly,  $s_{\xi_D(z)}^2-S_{\xi_D(z)}^2\xrightarrow{p}0$ and  $s_{\xi_Y(z)\xi_D(z)}-S_{\xi_Y(z)\xi_D(z)}\xrightarrow{p}0$. Thus,
	\begin{eqnarray}
	s_{A_{\cal}(z)}^2 & = & s_{\xi_{Y}(z)}^2 + \hat\tau_{\cal}^2 s_{\xi_D(z)}^2 - 2 \hat\tau_{\cal} s_{\xi_Y(z)\xi_D(z)} \nonumber \\
	&\xrightarrow{p} & \lim_{n\rightarrow\infty} \big\{ S_{\xi_{Y}(z)}^2 + \tau^2 S_{\xi_D(z)}^2 - 2 \tau S_{\xi_Y(z)\xi_D(z)} \big\} = \lim_{n\rightarrow\infty} S_{A_{\cal}(z)}^2. \nonumber
	\end{eqnarray}
	Therefore,
	$$
	\hat \sigma^2_{\cal} \xrightarrow{p}\lim_{n\rightarrow\infty}\frac{n}{\tau_D^2}\Big\{\frac{S_{A_{\cal}(1)}^2}{n_1}+\frac{S_{A_{\cal}(0)}^2}{n_0}\Big \} \geq \lim_{n\rightarrow\infty}\sigma^2_{\cal}.
	$$
	
\end{proof}

\subsection{Proof of Theorem \ref{thm::cal var}}

\begin{proof}
	According to Theorem~\ref{thm::cal}, the asymptotic variance of $n^{1/2}\hat\tau_{\cal}$ is
	$$
	\lim_{n\rightarrow\infty}\frac{n}{\tau_D^2}\Big\{\frac{S_{A_{\cal}(1)}^2}{n_1}+\frac{S_{A_{\cal}(0)}^2}{n_0}- \frac{S_{A_{\cal}(1)-A_{\cal}(0)}^2}{n}\Big\}.
	$$
	Note that, for $z=0,1$,
	$$
	\begin{aligned}
	A_{\cal,i}(z)&=Y_i(z)-\tau D_i(z)-(\W_i-\bar {\W})^{\T}\{\gamma_Y(z)-\tau\gamma_D(z)\}\\
	&=A_i(z)-(\W_i-\bar {\W})^{\T}\{\gamma_Y(z)-\tau\gamma_D(z)\},
	\end{aligned}
	$$
	where $A_i(z)=Y_i(z)-\tau D_i(z)$. Thus, similar to the proof of Theorem \ref{thm::ols var}, we have
	$$
	S_{A_{\cal}(1)}^2=S_{A(1)}^2-\{\gamma_Y(1)-\tau\gamma_D(1)\}^{\T}S_{\W}^2\{\gamma_Y(1)-\tau\gamma_D(1)\},
	$$
	$$
	S_{A_{\cal}(0)}^2=S_{A(0)}^2-\{\gamma_Y(0)-\tau\gamma_D(0)\}^{\T}S_{\W}^2\{\gamma_Y(0)-\tau\gamma_D(0)\}.
	$$
	Let $\phi(1)=\gamma_Y(1)-\tau\gamma_D(1)$ and  $\phi(0)=\gamma_Y(0)-\tau\gamma_D(0)$, then 
	$$
	S_{A_{\cal}(1)}^2=S_{A(1)}^2-\phi(1)^{\T}S_{\W}^2\phi(1),
	\quad 
	S_{A_{\cal}(0)}^2=S_{A(0)}^2-\phi(0)^{\T}S_{\W}^2\phi(0).
	$$
	As
	$$
	A_{\cal,i}(1)-A_{\cal,i}(0)=A_i(1)-A_i(0)-(\W_i-\bar {\W})^{\T}\{\phi(1)-\phi(0)\},
	$$
	we have
	$$
	S_{A_{\cal}(1)-A_{\cal}(0)}^2=S_{A(1)-A(0)}^2-\{\phi(1)-\phi(0)\}^{\T}S_{\W}^2\{\phi(1)-\phi(0)\}.
	$$
	Therefore, the difference between the asymptotic variances of $n^{1/2}\hat\tau_{\cal}$ and $n^{1/2}\hat\tau_{\wald}$ is the limit of $1/\tau_D^2$ times
	
	$$
	\begin{aligned}
	&-\frac{\phi(1)^{\T} S_{\W}^2 \phi(1)}{p_{1}}-\frac{\phi(0)^{\T}  S_{\W}^2  \phi(0)}{p_{0}}+\{\phi(1)-\phi(0)\}^{\T}  S_{\W}^2 \{\phi(1)-\phi(0)\} \\
	=&-\frac{1}{p_{1} p_{0}}[p_{0} \phi(1)^{\T} S_{\W}^2 \phi(1)+p_{1} \phi(0)^{\T} S_{\W}^2\phi(0)-p_{1} p_{0}\{\phi(1)-\phi(0)\}^{\T} S_{\W}^2\{\phi(1)-\phi(0)\}]\\
	=&-\frac{1}{p_{1} p_{0}}[p_{0}^{2} \phi(1)^{\T} S_{\W}^2\phi(1)+p_{1}^{2} \phi(0)^{\T} S_{\W}^2\phi(0)+2 p_{1} p_{0} \phi(1)^{\T} S_{\W}^2 \phi(0)] \\
	=&-\frac{1}{p_{1} p_{0}}\{p_{0} \phi(1)+p_{1} \phi(0)\}^{\T} S_{\W}^2\{p_{0} \phi(1)+p_{1} \phi(0)\} \\
	\leq& 0.
	\end{aligned}
	$$
	Moreover, since
	$$
	\hat \sigma^2_{\wald} \xrightarrow{p}\lim_{n\rightarrow\infty}\frac{n}{\tau_D^2}\Big\{\frac{S_{A(1)}^2}{n_1}+\frac{S_{A(0)}^2}{n_0}\Big \},
	\quad
	\hat \sigma^2_{\cal} \xrightarrow{p}\lim_{n\rightarrow\infty}\frac{n}{\tau_D^2}\Big\{\frac{S_{A_{\cal}(1)}^2}{n_1}+\frac{S_{A_{\cal}(0)}^2}{n_0}\Big \},
	$$
	and $n_z / n \rightarrow p_z$, the difference between the variance estimators $\hat\sigma^2_{\cal}$ and $\hat\sigma^2_{\wald}$ converges in probability to
	$$
	-\lim_{n\rightarrow\infty}\frac{1}{\tau_D^2}\Big\{\frac{\phi(1)^{\T} S_{\W}^2 \phi(1)}{p_1}+\frac{\phi(0)^{\T}  S_{\W}^2  \phi(0)}{p_0} \Big\}\leq 0.
	$$
	
	
	Next, we compare the asymptotic variances and variance estimators of $\hat\tau_{\cal}$ and $\hat\tau_{\logit}$. 
	During the proof of Theorems \ref{thm::logit} and \ref{thm::cal}, we have shown that $\hat\tau_{\logit,Y} -\tau \hat\tau_{\logit,D}$ differs by $o_p(n^{-1/2})$ from the following difference in means: 
	$$
	\frac{1}{n_1}\sum_{i:Z_i=1}\left\{\eta_{Y,i}(1)-\tau\eta_{D,i}(1)\right\}-\frac{1}{n_0}\sum_{i:Z_i=0}\left\{\eta_{Y,i}(0)-\tau\eta_{D,i}(0)\right\},
	$$
	and $\hat\tau_{\cal,Y} -\tau \hat\tau_{\cal,D}$ differs by $o_p(n^{-1/2})$ from the following difference in means: 
	$$
	\frac{1}{n_1}\sum_{i:Z_i=1}\left\{\xi_{Y,i}(1)-\tau\xi_{D,i}(1)\right\}-\frac{1}{n_0}\sum_{i:Z_i=0}\left\{\xi_{Y,i}(0)-\tau\xi_{D,i}(0)\right\}.
	$$
	Unpacking the notation, we have,
	$$
	\begin{aligned}
	\hat\tau_{\logit,Y} -\tau \hat\tau_{\logit,D}
	=&\frac{1}{n_1}\sum_{i:Z_i=1}\Big\{ Y_i(1)-\tau D_i(1)-\W_i^\T(1,0,-\tau,0)^\T \Big\} \\
	&-\frac{1}{n_0}\sum_{i:Z_i=0}\Big\{ Y_i(0)-\tau D_i(0)-W_i^\T(0,1,0,-\tau)^\T\Big\} +o_p(n^{-1/2}),
	\end{aligned}
	$$
	$$
	\begin{aligned}
	\hat\tau_{\cal,Y} -\tau \hat\tau_{\cal,D}
	=&\frac{1}{n_1}\sum_{i:Z_i=1}\Big[ Y_i(1)-\tau D_i(1)-\alpha_1- W_i^\T\big\{ \gamma_{Y}(1)-\tau \gamma_{D}(1) \big\} \Big]\\
	&-\frac{1}{n_0}\sum_{i:Z_i=0}\Big[ Y_i(0)-\tau D_i(0)-\alpha_0- W_i^\T\big\{ \gamma_{Y}(0)-\tau \gamma_{D}(0) \big\} \Big] +o_p(n^{-1/2}),
	\end{aligned}
	$$
	Both equations are, up to an $o_p(n^{-1/2})$ difference, of the form
	$$
	\begin{aligned}
	&\frac{1}{n_1}\sum_{i:Z_i=1}\Big\{Y_i(1)-\tau D_i(1)-a_1-\W_i^\T b_1\Big\}
	-\frac{1}{n_0}\sum_{i:Z_i=0}\left\{Y_i(0)-\tau D_i(0)-a_0-\W_i^\T b_0 \right\},
	\end{aligned}
	$$
	with the only difference being that in $\hat\tau_{\logit}$ we take $(a_1, b_1^{\T})^{\T}=(0,1,0,-\tau,0)^{\T}$  and $(a_0, b_0^{\T})^{\T} = (0, 0,1,0,-\tau)^{\T}$, whereas in $\hat\tau_{\cal}$ we take $(a_1, b_1^{\T})^{\T}=(\alpha_1,\{\gamma_Y(1)-\tau \gamma_D(1)\}^{\T})^{\T}$ and $(a_0, b_0^{\T})^{\T}=(\alpha_0,\{\gamma_Y(0)-\tau \gamma_D(0)\}^{\T})^{\T}$. Since the $o_p(n^{-1/2})$ term contributes nothing to the asymptotic variances of $\hat\tau_{\logit}$ and $\hat\tau_{\cal}$, we neglect it in the remainder of the proof.


	By the argument presented in Section 4.1 of \citet{lin2013}, the variance is minimized when $(a_1,b_1^{\T},a_0,b_0^{\T})^{\T}$ is taken to be the population OLS linear regression intercepts and slopes. This is exactly the case for $\hat\tau_{\cal}$, whereas $\hat\tau_{\logit}$ presents a feasible, but not necessarily optimal solution to the OLS problem. Consequently, the asymptotic variance of $\hat\tau_{\cal}$ does not exceed that of $\hat\tau_{\logit}$.

	According to Theorems~\ref{thm::logit} and \ref{thm::cal}, the probability limits of $\hat\sigma^2_{\logit}$ and $\hat\sigma^2_{\cal}$ are, respectively, 
	$$
	\lim_{n\rightarrow\infty}\frac{n}{\tau_{D}^2}\Big\{\frac{S_{A_{\logit}(1)}^2}{n_1}+\frac{S_{A_{\logit}(0)}^2}{n_0}\Big\}, \quad \lim_{n\rightarrow\infty}\frac{n}{\tau_D^2}\Big\{\frac{S_{A_{\cal}(1)}^2}{n_1}+\frac{S_{A_{\cal}(0)}^2}{n_0}\Big\}.
	$$
	Recall that, $\W_i=(\mu_{Y,i}(1), \mu_{Y,i}(0),\mu_{D,i}(1),\mu_{D,i}(0))^{\T}$,
	$$
	A_{\logit,i}(z)=Y_i(z)-\tau D_i(z)-\{\mu_{Y,i}(z)-\tau \mu_{D,i}(z)\},
	$$
	$$
	A_{\cal,i}(z)  = Y_i(z)-\tau D_i(z)-(\W_i-\bar{\W})^{\T}\{\gamma_Y(z)-\tau \gamma_D(z) \}.
	$$
	As $\gamma_Y(z)-\tau \gamma_D(z)$ is the  slop  of regressing $Y_i(z)-\tau D_i(z)$ on $\W_i$ with an intercept, we have $S_{A_{\cal}(z)}^2 \leq S_{A_{\logit}(z)}^2$, $z=0,1$. Thus, the probability limit of $\hat\sigma^2_{\cal}$ is less than or equal to that of $\hat\sigma^2_{\logit}$.
	
\end{proof}

\subsection{Proof of Theorems~\ref{thm::MCATE} and \ref{thm::MCATE var}}

\begin{proof}
	
	The proofs of Theorems~\ref{thm::MCATE}  and \ref{thm::MCATE var} are similar to that of the results on CATE, just replacing $Y$ and $D$ by $G$ and $H$, respectively.
	
\end{proof}

\section{Additional simulation results}

\begin{figure}
	\centering
	\includegraphics[width=0.9\linewidth]{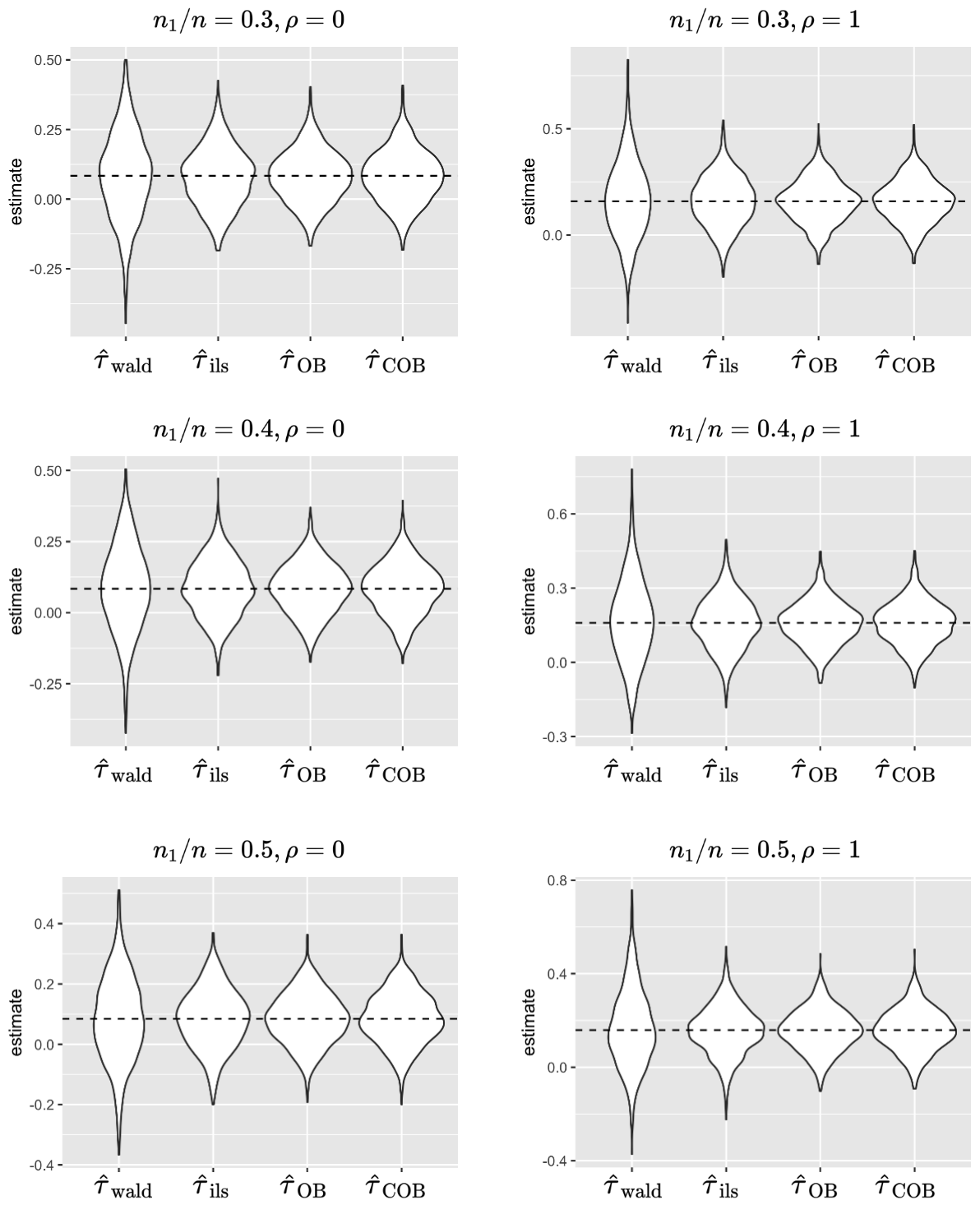}
	\caption{\label{fig::CATE200}Violin plot for the distributions of CATE estimators ($n=200$). The dotted line is the true value of the CATE.}
\end{figure}

\begin{figure}
	\centering
	\includegraphics[width=0.9\linewidth]{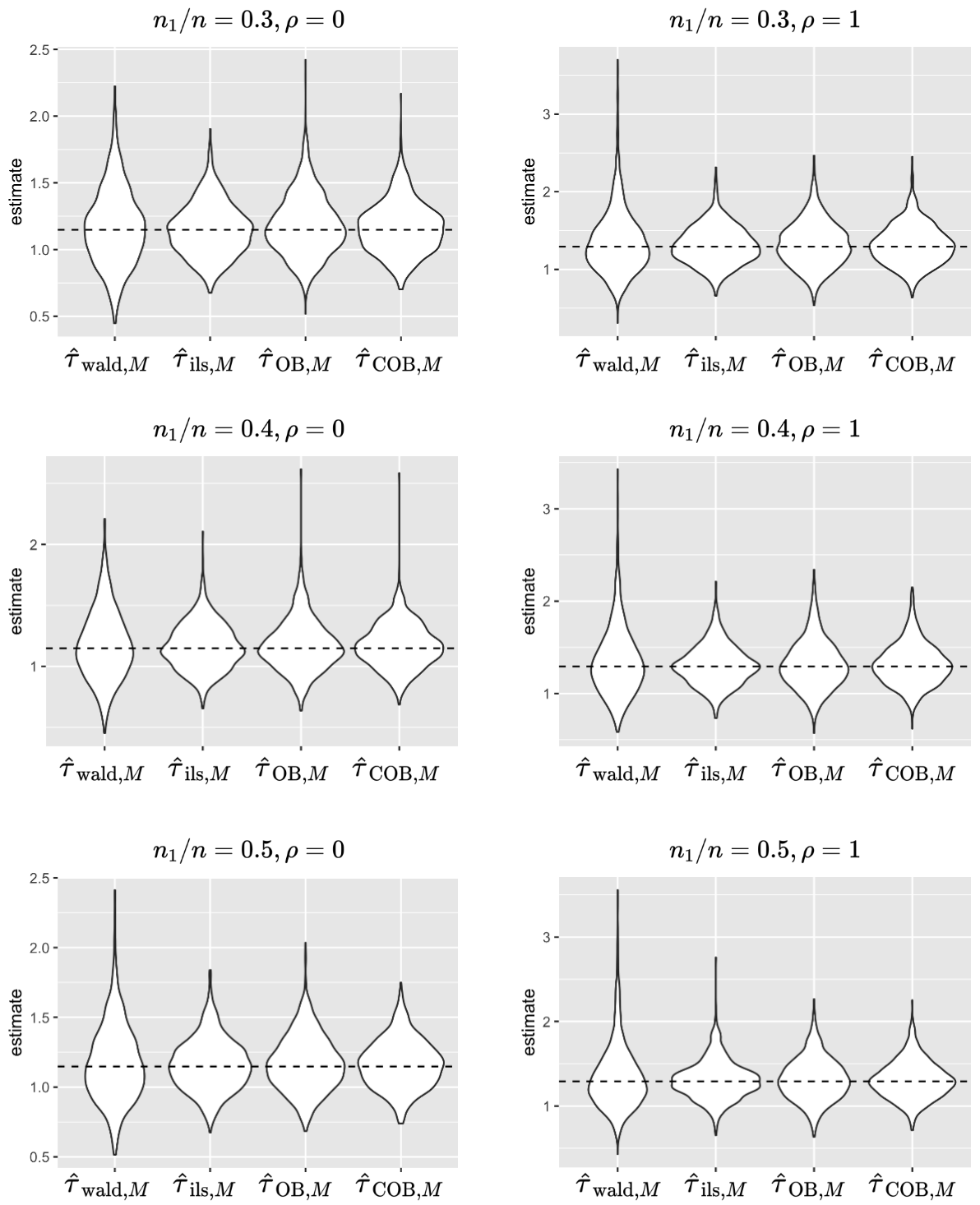}
	\caption{\label{fig::MCATE200}Violin plot for the distributions of MCATE estimators ($n=200$). The dotted line is the true value of the MCATE.}
\end{figure}

\begin{figure}
	\centering
	\includegraphics[width=0.9\linewidth]{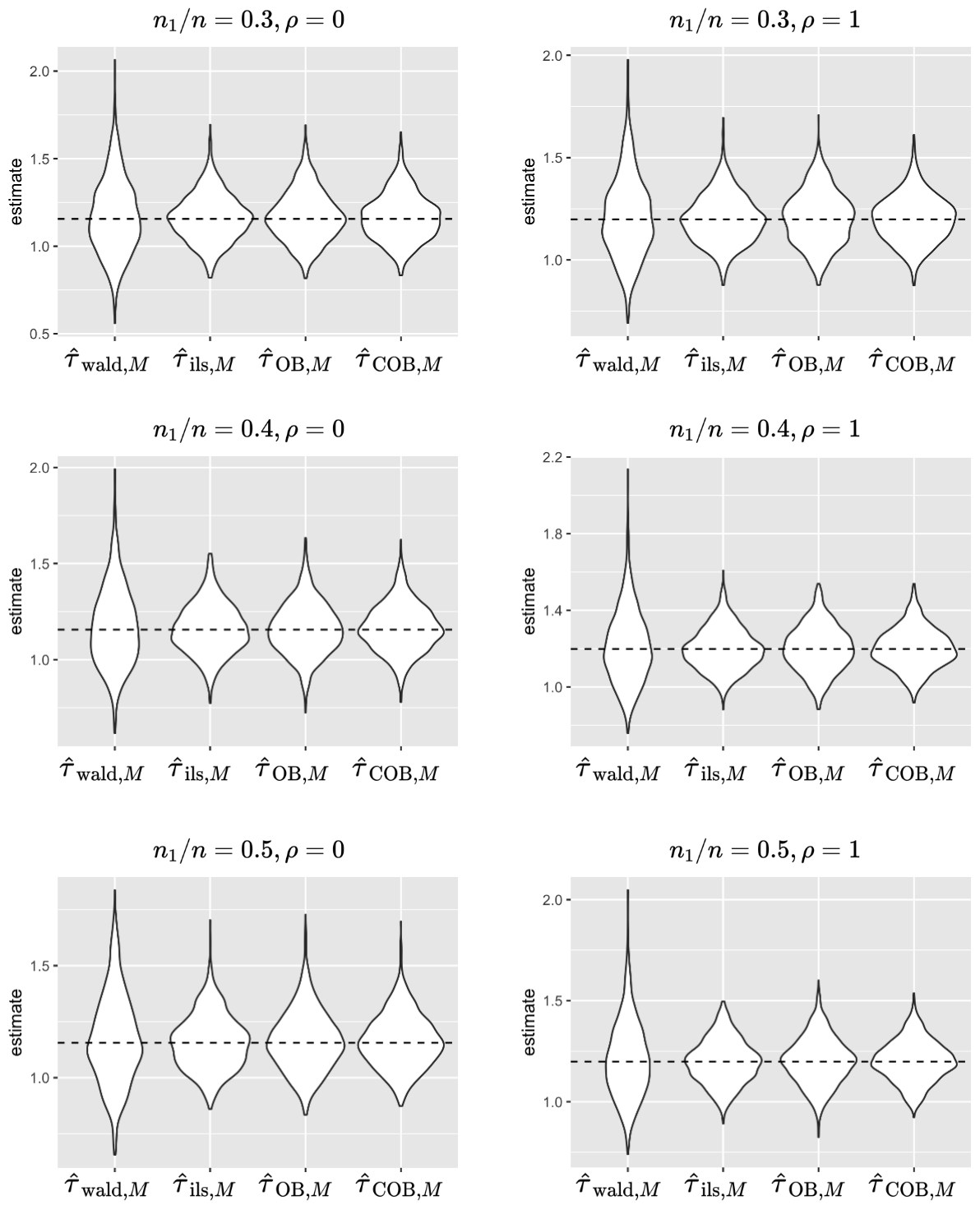}
	\caption{\label{fig::MCATE500}Violin plot for the distributions of MCATE estimators ($n=500$). The dotted line is the true value of MCATE.}
\end{figure}

In this section, we present the simulation results for estimating CATE ($n=200$) and MCATE; see Figures~\ref{fig::CATE200}-\ref{fig::MCATE500} and Tables~\ref{tab::CATE200}-\ref{tab::MCATE500}. The conclusions for the CATE estimators are similar to those in the main text. For MCATE, when $n=500$,  $\hat\tau_{\ils}$, $\hat\tau_{\logit}$, and $\hat\tau_{\cal}$ also perform better than $\hat\tau_{\wald}$, while the improvement of $\hat\tau_{\logit}$ over $\hat\tau_{\ils}$ is not as significant as the cases of CATE estimation, and $\hat\tau_{\cal}$ performs better than $\hat\tau_{\logit}$. When $n=200$, the MCATE estimators have some large values (looks like outliers; especially for the Wald estimator) in certain cases, mainly because of the estimations of $\tau_H$ are close to zero in some of the randomization realizations. Moreover, $\hat\tau_{\logit}$ does not work as well as $\hat\tau_{\ils}$, but the calibrated estimator $\hat\tau_{\cal}$ still works the best. 

\begin{table}
	\centering
	\caption{\label{tab::CATE200}Performance of CATE estimators ($n=200$)}
		\begin{threeparttable}
	\begin{tabular}{cccccccccc}
	\hline
	&\multirow{2}{*}{$\rho$}& \multirow{2}{*}{$n_1/n$} & \multirow{2}{*}{Bias} & \multirow{2}{*}{SD} & \multirow{2}{*}{RMSE} & RMSE & \multirow{2}{*}{CP} & CI & Length \\ 
	& & &  &  &  & ratio &  & length & ratio\\ 
	\hline	
	$\hat\tau_{\wald}$& 0 & 0.3  & -0.001 & 0.153 & 0.153 & 1.000 & 0.971 & 0.636 & 1.000 \\ 
	$\hat\tau_{\ils}$ & 0& 0.3  & -0.004 & 0.105 & 0.105 & 0.687 & 0.966 & 0.439 & 0.691 \\ 
	$\hat\tau_{\logit}$  & 0 & 0.3& -0.003 & 0.092 & 0.092 & 0.602 & 0.957 & 0.376 & 0.591 \\ 
	$\hat\tau_{\cal}$ & 0& 0.3  & -0.002 & 0.093 & 0.093 & 0.611 & 0.951 & 0.376 & 0.590 \\ 
	\hline
	$\hat\tau_{\wald}$ & 0& 0.4  & 0.001 & 0.149 & 0.149 & 1.000 & 0.954 & 0.594 & 1.000 \\ 
	$\hat\tau_{\ils}$ & 0& 0.4  & 0.001 & 0.099 & 0.099 & 0.664 & 0.967 & 0.412 & 0.694 \\ 
	$\hat\tau_{\logit}$ & 0& 0.4  & -0.000 & 0.087 & 0.087 & 0.585 & 0.962 & 0.355 & 0.598 \\ 
	$\hat\tau_{\cal}$& 0 & 0.4  & -0.000 & 0.089 & 0.089 & 0.595 & 0.953 & 0.355 & 0.598 \\ 
	\hline
	$\hat\tau_{\wald}$ & 0 & 0.5  & -0.002 & 0.142 & 0.142 & 1.000 & 0.960 & 0.583 & 1.000 \\ 
	$\hat\tau_{\ils}$  & 0 & 0.5 & -0.000 & 0.097 & 0.097 & 0.683 & 0.967 & 0.404 & 0.693 \\ 
	$\hat\tau_{\logit}$  & 0& 0.5  & 0.000 & 0.084 & 0.084 & 0.590 & 0.970 & 0.349 & 0.599 \\ 
	$\hat\tau_{\cal}$ & 0 & 0.5  & -0.000 & 0.084 & 0.084 & 0.595 & 0.965 & 0.349 & 0.599 \\ 
	\hline
	$\hat\tau_{\wald}$ & 1& 0.3  & 0.008 & 0.178 & 0.178 & 1.000 & 0.964 & 0.729 & 1.000 \\ 
	$\hat\tau_{\ils}$ & 1& 0.3  & -0.000 & 0.117 & 0.117 & 0.660 & 0.962 & 0.487 & 0.668 \\ 
	$\hat\tau_{\logit}$& 1 & 0.3  & -0.001 & 0.100 & 0.100 & 0.563 & 0.960 & 0.415 & 0.569 \\ 
	$\hat\tau_{\cal}$ & 1 &0.3 & -0.000 & 0.100 & 0.100 & 0.565 & 0.964 & 0.414 & 0.568 \\ 
	\hline
	$\hat\tau_{\wald}$& 1  &0.4 & 0.010 & 0.165 & 0.165 & 1.000 & 0.962 & 0.676 & 1.000 \\ 
	$\hat\tau_{\ils}$& 1  &0.4 & -0.000 & 0.104 & 0.104 & 0.629 & 0.972 & 0.455 & 0.672 \\ 
	$\hat\tau_{\logit}$ & 1& 0.4  & -0.001 & 0.086 & 0.086 & 0.520 & 0.968 & 0.390 & 0.577 \\ 
	$\hat\tau_{\cal}$ & 1 & 0.4 & -0.001 & 0.087 & 0.087 & 0.525 & 0.968 & 0.389 & 0.575 \\ 
	\hline
	$\hat\tau_{\wald}$& 1 &0.5 & 0.007 & 0.164 & 0.165 & 1.000 & 0.969 & 0.663 & 1.000 \\ 
	$\hat\tau_{\ils}$& 1 & 0.5  & -0.002 & 0.104 & 0.104 & 0.631 & 0.963 & 0.445 & 0.672 \\ 
	$\hat\tau_{\logit}$ & 1 & 0.5 & -0.001 & 0.089 & 0.089 & 0.539 & 0.967 & 0.383 & 0.578 \\ 
	$\hat\tau_{\cal}$ & 1 &0.5 & -0.001 & 0.089 & 0.089 & 0.542 & 0.961 & 0.382 & 0.576 \\ 
	\hline
\end{tabular}
	
	Note: SD, standard deviation; RMSE, root of mean squared error; RMSE ratio, relative to the Wald  estimator; CP, empirical coverage probability of the $95\%$ confidence intervals; CI length, mean confidence interval length of the $95\%$ confidence intervals; Length ratio, relative to the Wald estimator.
			\end{threeparttable}
\end{table}

\begin{table}
	\centering
	\caption{\label{tab::MCATE200}Performance of MCATE estimators  ($n=200$)}
		\begin{threeparttable}
	\begin{tabular}{cccccccccc}
	\hline
	&\multirow{2}{*}{$\rho$}& \multirow{2}{*}{$n_1/n$} & \multirow{2}{*}{Bias} & \multirow{2}{*}{SD} & \multirow{2}{*}{RMSE} & RMSE & \multirow{2}{*}{CP} & CI & Length \\ 
	& & &  &  &  & ratio &  & length & ratio\\ 
	\hline	
	$\hat\tau_{\wald}$  & 0& 0.3& 0.020 & 0.287 & 0.287 & 1.000 & 0.967 & 1.212 & 1.000 \\ 
	$\hat\tau_{\ils}$ & 0 & 0.3 & 0.003 & 0.198 & 0.198 & 0.689 & 0.969 & 0.824 & 0.680 \\ 
	$\hat\tau_{\logit}$ & 0 & 0.3 & 0.036 & 0.228 & 0.231 & 0.805 & 0.970 & 0.917 & 0.757 \\ 
	$\hat\tau_{\cal}$ & 0 & 0.3 & 0.016 & 0.194 & 0.195 & 0.678 & 0.975 & 0.781 & 0.645 \\ 
	\hline
	$\hat\tau_{\wald}$ & 0& 0.4  & 0.027 & 0.282 & 0.284 & 1.000 & 0.957 & 1.152 & 1.000 \\ 
	$\hat\tau_{\ils}$ & 0 &0.4 &  0.014 & 0.188 & 0.189 & 0.665 & 0.974 & 0.790 & 0.685 \\ 
	$\hat\tau_{\logit}$  & 0& 0.4 & 0.024 & 0.210 & 0.212 & 0.746 & 0.971 & 0.873 & 0.758 \\ 
	$\hat\tau_{\cal}$  & 0& 0.4 & 0.019 & 0.181 & 0.182 & 0.643 & 0.966 & 0.739 & 0.642 \\ 
	\hline
	$\hat\tau_{\wald}$ & 0 & 0.5 & 0.023 & 0.274 & 0.275 & 1.000 & 0.963 & 1.132 & 1.000 \\ 
	$\hat\tau_{\ils}$ & 0 & 0.5 & 0.012 & 0.185 & 0.186 & 0.675 & 0.965 & 0.775 & 0.685 \\ 
	$\hat\tau_{\logit}$ & 0& 0.5  & 0.022 & 0.198 & 0.200 & 0.726 & 0.969 & 0.862 & 0.762 \\ 
	$\hat\tau_{\cal}$ & 0 & 0.5 & 0.018 & 0.174 & 0.175 & 0.636 & 0.969 & 0.722 & 0.638 \\ 
	\hline
	$\hat\tau_{\wald}$ & 1& 0.3  & 0.066 & 0.419 & 0.425 & 1.000 & 0.955 & 1.685 & 1.000 \\ 
	$\hat\tau_{\ils}$ & 1 & 0.3 & 0.019 & 0.245 & 0.246 & 0.579 & 0.968 & 1.029 & 0.611 \\ 
	$\hat\tau_{\logit}$ & 1 & 0.3 & 0.036 & 0.291 & 0.293 & 0.690 & 0.964 & 1.193 & 0.708 \\ 
	$\hat\tau_{\cal}$ & 1 & 0.3 & 0.024 & 0.254 & 0.255 & 0.602 & 0.954 & 1.012 & 0.601 \\ 
	\hline
	$\hat\tau_{\wald}$  & 1& 0.4 & 0.060 & 0.379 & 0.383 & 1.000 & 0.946 & 1.536 & 1.000 \\ 
	$\hat\tau_{\ils}$ & 1 & 0.4 & 0.015 & 0.217 & 0.218 & 0.568 & 0.976 & 0.960 & 0.625 \\ 
	$\hat\tau_{\logit}$& 1  & 0.4 & 0.026 & 0.267 & 0.268 & 0.700 & 0.965 & 1.098 & 0.715 \\ 
	$\hat\tau_{\cal}$ & 1 & 0.4 & 0.024 & 0.225 & 0.226 & 0.589 & 0.968 & 0.948 & 0.617 \\ 
	\hline
	$\hat\tau_{\wald}$  & 1 & 0.5& 0.067 & 0.407 & 0.412 & 1.000 & 0.959 & 1.556 & 1.000 \\ 
	$\hat\tau_{\ils}$ & 1 & 0.5 & 0.016 & 0.227 & 0.228 & 0.552 & 0.968 & 0.963 & 0.619 \\ 
	$\hat\tau_{\logit}$ & 1& 0.5  & 0.025 & 0.250 & 0.251 & 0.609 & 0.968 & 1.075 & 0.691 \\ 
	$\hat\tau_{\cal}$ & 1 & 0.5 & 0.017 & 0.221 & 0.222 & 0.538 & 0.968 & 0.935 & 0.601 \\ 
	\hline
\end{tabular}

	Note: SD, standard deviation; RMSE, root of mean squared error; RMSE ratio, relative to the Wald  estimator; CP, empirical coverage probability of the $95\%$ confidence intervals; CI length, mean confidence interval length of the $95\%$ confidence intervals; Length ratio, relative to the Wald estimator.
		\end{threeparttable}
\end{table}

\begin{table}
	\centering
	\caption{\label{tab::MCATE500}Performance of MCATE estimators  ($n=500$)}
	\begin{threeparttable}
		\begin{tabular}{cccccccccc}
	\hline
	&\multirow{2}{*}{$\rho$}& \multirow{2}{*}{$n_1/n$} & \multirow{2}{*}{Bias} & \multirow{2}{*}{SD} & \multirow{2}{*}{RMSE} & RMSE & \multirow{2}{*}{CP} & CI & Length \\ 
	& & &  &  &  & ratio &  & length & ratio\\ 
	\hline	
	$\hat\tau_{\wald}$& 0 & 0.3  & 0.013 & 0.211 & 0.212 & 1.000 & 0.964 & 0.843 & 1.000 \\  
	$\hat\tau_{\ils}$ & 0& 0.3  & 0.007 & 0.134 & 0.134 & 0.635 & 0.966 & 0.563 & 0.668 \\ 
	$\hat\tau_{\logit}$ & 0& 0.3  & 0.016 & 0.143 & 0.144 & 0.680 & 0.975 & 0.616 & 0.731 \\ 
	$\hat\tau_{\cal}$ & 0& 0.3  & 0.011 & 0.132 & 0.133 & 0.628 & 0.966 & 0.548 & 0.650 \\ 
	\hline
	$\hat\tau_{\wald}$ & 0 & 0.4 & 0.014 & 0.205 & 0.205 & 1.000 & 0.961 & 0.796 & 1.000 \\ 
	$\hat\tau_{\ils}$ & 0& 0.4  & 0.005 & 0.130 & 0.130 & 0.632 & 0.960 & 0.532 & 0.668 \\ 
	$\hat\tau_{\logit}$ & 0 & 0.4 & 0.005 & 0.135 & 0.135 & 0.660 & 0.964 & 0.580 & 0.728 \\  
	$\hat\tau_{\cal}$ & 0 & 0.4 & 0.009 & 0.123 & 0.124 & 0.603 & 0.967 & 0.514 & 0.646 \\ 
	\hline
	$\hat\tau_{\wald}$ & 0 & 0.5 & 0.015 & 0.197 & 0.198 & 1.000 & 0.960 & 0.789 & 1.000 \\ 
	$\hat\tau_{\ils}$  & 0 & 0.5& 0.003 & 0.123 & 0.123 & 0.622 & 0.970 & 0.524 & 0.664 \\ 
	$\hat\tau_{\logit}$ & 0& 0.5  & 0.007 & 0.134 & 0.135 & 0.680 & 0.971 & 0.577 & 0.731 \\ 
	$\hat\tau_{\cal}$  & 0& 0.5 & 0.007 & 0.120 & 0.120 & 0.606 & 0.972 & 0.505 & 0.639 \\ 
	\hline
	$\hat\tau_{\wald}$ & 1 & 0.3 & 0.011 & 0.198 & 0.198 & 1.000 & 0.962 & 0.813 & 1.000 \\  
	$\hat\tau_{\ils}$& 1  & 0.3 & 0.002 & 0.119 & 0.119 & 0.600 & 0.971 & 0.514 & 0.632 \\ 
	$\hat\tau_{\logit}$ & 1 & 0.3 & 0.008 & 0.131 & 0.131 & 0.662 & 0.986 & 0.586 & 0.721 \\  
	$\hat\tau_{\cal}$ & 1& 0.3  & 0.002 & 0.113 & 0.113 & 0.572 & 0.968 & 0.490 & 0.603 \\ 
	\hline
	$\hat\tau_{\wald}$ & 1 & 0.4 & 0.013 & 0.190 & 0.190 & 1.000 & 0.957 & 0.770 & 1.000 \\ 
	$\hat\tau_{\ils}$ & 1 & 0.4 & 0.000 & 0.109 & 0.109 & 0.572 & 0.981 & 0.486 & 0.632 \\  
	$\hat\tau_{\logit}$ & 1 & 0.4 & 0.001 & 0.120 & 0.120 & 0.628 & 0.979 & 0.546 & 0.710 \\   
	$\hat\tau_{\cal}$ & 1 & 0.4 & 0.002 & 0.105 & 0.105 & 0.549 & 0.974 & 0.463 & 0.601 \\ 
	\hline
	$\hat\tau_{\wald}$  & 1 & 0.5& 0.011 & 0.188 & 0.189 & 1.000 & 0.957 & 0.761 & 1.000 \\ 
	$\hat\tau_{\ils}$ & 1 & 0.5 & -0.000 & 0.109 & 0.109 & 0.575 & 0.977 & 0.482 & 0.633 \\  
	$\hat\tau_{\logit}$ & 1 & 0.5 & 0.003 & 0.117 & 0.117 & 0.620 & 0.980 & 0.536 & 0.704 \\ 
	$\hat\tau_{\cal}$ & 1 & 0.5 & 0.002 & 0.101 & 0.101 & 0.535 & 0.975 & 0.456 & 0.600 \\ 
	\hline
\end{tabular}

		Note: SD, standard deviation; RMSE, root of mean squared error; RMSE ratio, relative to the Wald  estimator; CP, empirical coverage probability of the $95\%$ confidence intervals; CI length, mean confidence interval length of the $95\%$ confidence intervals; Length ratio, relative to the Wald estimator. 
		
	\end{threeparttable}
\end{table}

\end{document}
